\documentclass{aa}  
\usepackage{graphicx}
\usepackage{txfonts}
\usepackage{amsmath}
\begin{document} 
   \title{Refractory carbon depletion by photolysis through dust collisions and vertical mixing}

   \author{Vignesh Vaikundaraman \inst{1,2}, 
          Joanna Dr{\k{a}}{\.z}kowska\inst{1,2}, 
          Fabian Binkert \inst{2,3}, 
          Til Birnstiel \inst{2,3}
          \and
          Anna Miotello \inst{4}}

   \institute{Max Planck Institute for Solar System Research, Justus-von-Liebig-Weg 3, 37077 Göttingen, Germany \and University Observatory, Faculty of Physics, Ludwig-Maximilians-Universität München, Scheinerstr. 1, 81679 Munich, Germany \and
   Exzellenzcluster ORIGINS, Boltzmannstr. 2, D-85748 Garching, Germany \and
   European Southern Observatory, Karl-Schwarzschild-Strasse 2, 85748 Garching bei München, Germany}

   \date{Received XXXX; accepted YYYY}

    \abstract{The inner Solar System is depleted in refractory carbon in comparison to the interstellar medium and the depletion likely took place in the protoplanetary disk phase of the Solar System.}{We study the effect of photolysis of refractory carbon in the upper layers of the protosolar disk and its interplay with dust collisional growth and vertical mixing.}{We make use of a 1D Monte Carlo model to simulate dust coagulation and vertical mixing. To model the FUV flux of the disk, we use a simple analytical prescription and benchmark it with data from a radiative transfer simulation. We study the effects of fragmentation and bouncing on dust distribution and the propagation of carbon depletion.}{We find that when bouncing is included, the size distribution is truncated at smaller sizes than fragmentation-limited size distributions but there is a loss of small grains as well. The population of small grains is reduced due to fewer fragmentation events and this reduces the effectiveness of photolysis. We find that dust collisional growth and vertical mixing increase the effectiveness of carbon depletion by efficiently replenishing carbon to the upper regions of the disk with higher FUV flux. It takes around 100-300 kyr to reach the measured carbon abundances at 1 au, depending on the strength of the turbulence in the disk. These timescales are faster than reported by previous studies.}{Collisional redistribution and turbulent mixing are important aspects of dust evolution that should be included when modeling dust chemistry as they can influence the efficiency of chemical processes. Photolysis, along with another process such as sublimation, most likely played a key role in refractory carbon depletion that we see around us in the inner Solar System. }
 
   \keywords{planets and satellites: formation -- protoplanetary disks -- astrochemistry  -- methods: numerical -- planets and satellites: composition }
   \authorrunning{Vaikundaraman et al.}
   \titlerunning{Destruction of carbon in protoplanetary disks}
   \maketitle
\section{Introduction}\label{sec:solarsystemcarb}
 Our life on Earth as we know it is based on carbon. It is of prime importance that we get more insights into its abundance already during the protoplanetary disk/planetesimal stages as it plays an important role in setting the chemistry for life \citep{Krijt2023}. It is indeed ironic when observations point out that carbon-based life-bearing Earth and the inner Solar System are severely depleted in carbon in comparison with the interstellar medium \citep[ISM,][]{Bergin2015}. The carbon to silicon ratio for the Earth is lower than the Sun's carbon to silicon ratio by a factor of $10^{-4}$, whereas the outer solar system objects like some comets are not carbon depleted at all \citep{Bergin2015}. The reason for such a deficit in its carbon composition is still a puzzle. The composition of the material in planets and meteorites that we can observe and measure, is set by several disk processes and also by meteorite parent-body processes like hydrothermal activity \citep{2022GeCoA.318...83S, Lichtenberg}. Starting from the disk formation from the ISM material to dust evolution and planet formation, every process contributes to determining the composition of the dust. Meteorites are chunks from asteroids that have formed from primordial planetesimals that did not become planets. This means the meteorites, whose compositions we measure, are excellent ``fossils'' to understand the effects of different disk processes and consequently, the theory of planet formation. \citet{Bergin2015} showed that not all rocky objects in the Solar System show a carbon deficit. For example, the carbon abundance of Halley's comet is very similar to the ISM. \citet{Alexander2017} showed that the interstellar material is still retained in some Solar System objects. It is the inner Solar System or the terrestrial planet-forming region that has significant carbon depletion. 

Numerous studies have been conducted to measure the abundance of carbon in the solar system. About 50\% of the carbon is locked in the refractory form \citep{Zubko2004}. Furthermore, in a comprehensive study of primordial materials presented by \citet{Gail2017}, they found that 60\% of the carbon in the solid phase is found in the complex organic matter along with hydrogen, nitrogen, oxygen, and sulfur, 10\% is found as pure carbon and the rest are found in aromatic, aliphatic compounds and other components. A clear trend that can be noticed is that the meteorites that formed relatively early e.g., LL, L chondrite parent bodies are more depleted in carbon than the ones formed later, like the CI, CM chondrites \citep{McCubbin2019, Gail2017}. The trend is even more evident when we look at how little carbon is present in non-carbonaceous iron meteorites which form much earlier \citep{Goldstein2017, Kruijer2017}. This points to a carbon depletion process in the earlier stages of planet formation when the material was still in the form of dust. Such an early process leaves a remarkable footprint on the composition of planets. Measuring the carbon depletion on Earth is a big clue to the process of planet formation but is very challenging. The carbon content present in the core is still being estimated and a lot of models have been developed. A study by \citet{Fischer2020} estimated the core to be the largest carbon reservoir of the Earth. Even though the core is the largest carbon inventory, in absolute terms, carbon is present at 0.09--0.4\% by weight, which reaffirms how depleted the Earth is in carbon. A detailed summary of the carbon reservoir in the Earth and the Solar System is presented in \citet{Binkert2023}. 

Recent planetesimal formation models arrive at estimates that the earliest planetesimals form within 1 Myr \citep{Drazkowska2018, Lichtenberg2021, Morbidelli2022}. In slightly different contexts, pressure bumps can rapidly produce planetary cores, within 0.1 Myr \citep{Lau2022}, and studies on planet formation history around polluted white dwarfs also point to rapid planetesimal formation \citep{Bonsor2022}. All these studies necessitate a rapid carbon depletion process. \citet{Johansen2021} proposed that hot envelopes are created when pebbles are accreted onto the protoplanet leading to the sublimation of refractories. This is one of the possible reasons that Earth could have a bulk of its carbon reservoir in its core but it does not explain the deficiency of carbon in early-formed meteorites. So we focus on the earlier processes, that is the processes leading to planetesimal formation as they contribute majorly to the carbon content of meteorites. This means that the time to process refractory carbon by whatever means required should be before the parent bodies differentiation i.e. $\sim 1$Myr. 
In this paper, we focus on photolysis, an ultra-violet (UV) photon-induced depletion process (more details in Sec. \ref{sec:photolysis}). We look at the final factor of timing the process, the evolution of the far ultra-violet (FUV) luminosity. Models of FUV evolution by \citet{Kunitomo2021a} tell us that the FUV flux magnitudes do not fluctuate very much within the first Myr (or a few hundred kyr as an ambitious timescale) which is our estimated target depletion timescale. Several possible mechanisms have been proposed to tackle the puzzle of carbon depletion, which we will discuss in the remainder of the introduction.

\subsection{Depletion processes}\label{sec:photolysis}
\citet{Lee2010} proposed a scenario where the carbon is oxidized in the upper layers of the disks where it is hot enough. A study by \citet{Gail2017} suggested that oxidation was not enough to deplete the carbon in the inner disk. The other leading mechanisms still in the discussion are flash heating events, sublimation, and photolysis. Flash heating events were studied by \citet{Gail2017} where they qualitatively discussed short-term flash heating events in the disk. More recently, \citet{Li2021} proposed sublimation of carbon to be the primary process that leads to the depletion of carbon to arrive at observed abundances. \citet{Binkert2023} combined the effects of photolysis and sublimation to explore carbon depletion and transport with semi-analytical models and found that the combination of the processes could work only under specific conditions. In this paper, we look at the carbon depletion efficiency of photolysis when a proper dust collisional growth model is included.

Photolysis is the process of a breakdown of a chemical bond due to interaction with an ultraviolet (UV) photon. In the context of refractory carbon, it results in the formation and sublimation of hydrocarbons. In this paper, photolysis is used as a term to describe the whole sequence of events \citep{Alata2014, Alata2015}. Previously, photolysis of carbon was studied by \citet{Anderson2017} and \citet{Klarmann2018} and also by \citet{Binkert2023}. A common result of the analytical models is that there are two regimes of photolysis in a disk, the first one being unrestricted photolysis where the process of photolysis is not hampered by disk mechanisms such as transport or turbulence. In such a regime, photolysis is very efficient in depleting carbon if the vertical mixing is very efficient in bringing the carbon to the surface. The second regime is the residence time-limited photolysis. UV flux does not penetrate all the way through the disk. Consequently, only the upper layers of the disk are exposed to UV rays. We will call this region the exposed layer and the time the particles stay at the exposed layer is called the residence time. \citet{Klarmann2018} concluded that photolysis is ineffective when the effects of radial and vertical dust transport are included. They found that when vertical mixing processes are not efficient enough, particles stay longer in the exposed layer, leading to high residence times and slow depletion of carbon as already depleted grains stay longer in the exposed layer and are not mixed with carbon-rich grains. This leads to an inefficient depletion of carbon due to the depleted grains staying in the upper layers and shielding the UV photons from the carbon-rich grains. Radial transport makes depletion even more difficult and \citet{Klarmann2018} find that radial transport can even distribute carbon across the disk. Further studies by \citet{Binkert2023} found that vertical transport limits depletion more than radial transport and these effects can be overcome only after including pyrolysis of carbon and outburst events in their disk model. However, all these studies included (semi-)analytical models for dust evolution. In reality, a potential form of dust transport is the redistribution of carbonaceous material due to dust collisional growth. These effects are not included in these models, the primary reason being that analytical models do not cover the true picture of dust evolution owing to its very stochastic nature.

This paper is an alternative study to \citet{Binkert2023} who, building on \citet{Klarmann2018}, developed semi-analytical models to look at (refractory) carbon depletion timescales and examine photolysis and pyrolysis as a solution to the carbon deficit puzzle. We make use of a Monte Carlo dust evolution code (more details in Sect.~\ref{sec:1DMC}) to accurately describe dust evolution models and make use of an analytical model that includes scattering effects (Sec. \ref{subsecFUV-analytical}) which we benchmark against a sophisticated FUV flux model (Sec. \ref{subsec-DALI}). We finally look at depletion timescales (Sec. \ref{sec:results}) and compare it with the estimates from \citet{Binkert2023} and examine the viability of photolysis as a solution to the puzzle in the presence of proper dust evolution models. 

\section{Methods}\label{sec:1DMC}

\begin{figure*} [ht]
    \centering
    \includegraphics[scale=0.44]{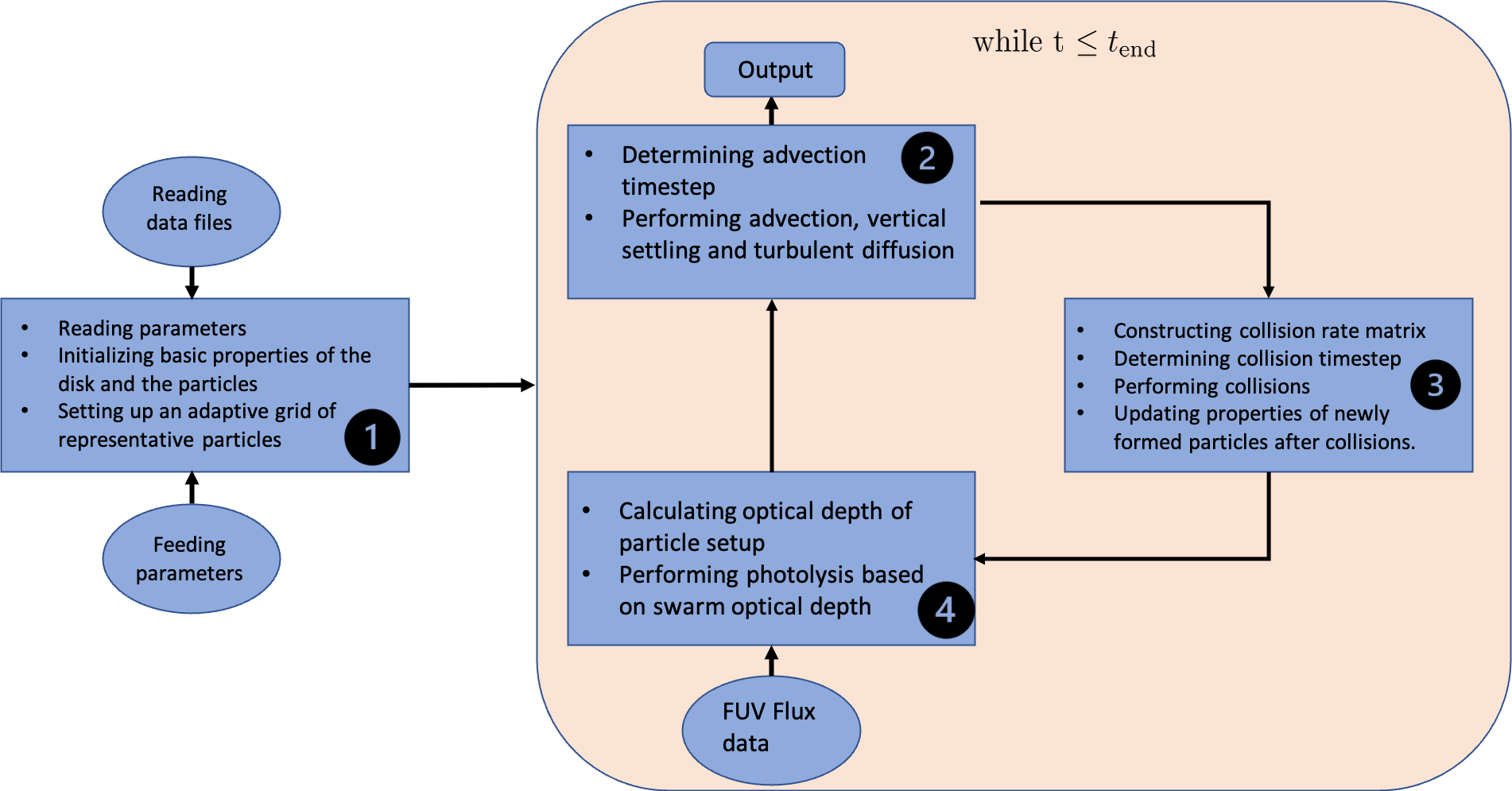}
    \caption[1DMC code workflow]{The workflow of the Monte Carlo code we use.}
    \label{fig:Workflow}
\end{figure*}

We use the Monte Carlo dust coagulation code presented in \citet{Drazkowska2013}. The code is written in \texttt{FORTRAN 90} and parallelized using \texttt{OpenMP}. We model a one-dimensional vertical disk column placed at a distance of 1 AU from a solar-mass star. We include dust collisional growth and fragmentation, vertical settling, turbulent mixing, and photolysis. An analytical prescription is used to describe the gas disk structure and we use the representative particle approach to treat dust coagulation. Instead of following every physical particle, we track a limited number of representative particles and the collisions they go through. The fundamental working of the dust evolution aspects of the code (settling, mixing, collisions) is the same as in \citet{Drazkowska2013}. The only difference is that we implement bouncing as an additional collisional outcome and add the chemistry of photolysis to the code. The workflow of the code is presented in Fig.~\ref{fig:Workflow}. We implement bouncing as an elastic collision in which the properties of the representative particle after a bouncing event are unchanged \citep{Guttler2010}. We do not include porosity evolution as done by \citet{Zsom2010} where bouncing would then act as a compaction process for the particles. The collisional outcomes are decided by the relative velocities of the particles as a piece-wise function,

\begin{equation}
\text{collisional outcome} = 
     \begin{cases}
       \text{sticking}  & \Delta V_{ij} < v_{\text{bounce}}, \\
       \text{bouncing}   & v_{\text{bounce}} \le \Delta V_{ij} $<$ v_{\text{frag}}, \\
      \text{fragmentation} &  \Delta V_{ij} \ge v_{\text{frag}},\\ 
    \end{cases}    
\end{equation}
where $\Delta V_{ij}$ is the relative velocity between the representative particle $i$ and the physical particle (represented by the representative particle) $j$. We also implement a small probability P = $10^{-4}$ that particles fragment in the bouncing regime i.e. when their velocities are below the fragmentation velocity $v_{\text{frag}}$. More discussion on the fragmentation and bouncing barriers is in Sect.~\ref{Sec-simtest}.

\subsection{Photolysis implementation}
To track the evolution of carbon abundances of the particles, we extended a few aspects of the code. One such new aspect is the effect of photolysis. In this work, photolysis is defined as the release of hydrocarbons from the dust grain surface due to the interaction of the dust grains with the FUV photons. The rate of loss is given by \citep{Binkert2023}
\begin{equation}
    R_{\mathrm{FUV}} = \sigma Y_{\mathrm{ph}} F_{\mathrm{FUV}},
\end{equation}
where $R_{\mathrm{FUV}}$ is the rate of carbon atoms lost per unit of time, $\sigma = \pi a^2$ is the geometrical cross-section of the grain, $Y_{\mathrm{ph}} = 8 \times 10^{-4}$ is the yield of photolysis per photon derived from experiments \citep{Alata2014, Alata2015}, and $F_{\mathrm{FUV}}$ is the stellar flux integrated over the FUV wavelengths ([$F_{\mathrm{FUV}}$] = cm$^{-2}$s$^{-1}$) at the optical depth at which the grain is present. To find the mass loss rate of carbon we calculate
\begin{equation}
    \Dot{M_c} = m_c R_{\mathrm{FUV}},
\end{equation}
where $m_c$ is the mass of a carbon atom. For a given timestep, the loss in mass of carbon in a grain can be calculated as,
\begin{equation}\label{eq:photolysis}
    \Delta M_c = m_c \pi a^2 Y_{\mathrm{ph}} F_{\mathrm{FUV}} \Delta t,
\end{equation}
where $\Delta t$ is the timestep duration. From Eq. \ref{eq:photolysis} above, it is clear that there are two ingredients here that we need to simulate photolysis: dust and FUV photons. The former is well established in the code setup but the latter is not. In the following section we describe the implementation of the FUV model.

\subsubsection{FUV flux}\label{Sec-FUV}
In our code, we take in the FUV flux as input. The input file contains the FUV flux $F_{\mathrm{FUV,input}}$ as a function of optical depth $\tau_{\mathrm{input}}$. We then calculate the optical depth of the simulated dust column $\tau_{\mathrm{dust}}$ and use an interpolation routine to arrive at $F_{\mathrm{FUV,dust}}$ using $F_{\mathrm{FUV,input}}$, $\tau_{\mathrm{input}}$ and $\tau_{\mathrm{dust}}$. There have been simple prescriptions for the FUV flux used e.g. \citet{Klarmann2018} assumed that the FUV flux has a hard boundary at optical depth $\tau=1$ ($z\approx 3H$), below which FUV flux is completely extinct. Similarly, in the context of sugar synthesis on icy dust particles, \citet{Takehara2022a} modeled the UV flux where they had a boundary at $z = 2.5H$, i.e. the particles below this height are completely shielded from the UV rays. Even though \citet{VanZadelhoff2003} showed that FUV photons do penetrate deeper ($\tau > 1$) into the disk, mapping the far ultra-violet flux in protoplanetary disks has always been a challenge. Photon propagation and scattering go hand in hand and more so in the case of protoplanetary disks as we are looking at regions with higher dust densities. In the context of carbon depletion in protoplanetary disks, \citet{Anderson2017} used a sophisticated Monte Carlo radiative transfer model proposed by \citet{Bethell2011} to arrive at the FUV and X-ray flux. Similarly, \citet{Clepper2022} tackled FUV propagation in combination with chemistry and dust evolution in the context of CO depletion in protoplanetary disks. 

We build an analytical model that takes into account single scattering events of FUV photons by the dust grains and resonant scattering of Lyman-$\alpha$ (Ly-$\alpha$) photons by the H-atoms. In order to benchmark the model and compare it with proper simulations, we make use of the radiative transfer code \texttt{DALI} to get the FUV flux data as a function of optical depth at 1 AU. The details of the analytical model and the benchmarking with the simulation are discussed in Sect. \ref{subsecFUV-analytical}.

\section{Model}\label{Sec-simtest}
We have outlined the workings of the code and the FUV flux setup. In this section, we explore the model and the outcomes to bring forward the physics of dust evolution before photolysis is brought into the picture.
\subsection{Disk model}
For modeling the protosolar disk, we take a $M_\star = M_\odot$, $L_\star = L_\odot$ and $L_{\mathrm{UV}} = 0.01L_\star$ \citep{Sokal2018,Siebenmorgen2010}. The  FUV luminosity of classical T Tauri stars is proportional to its mass accretion rate as well 
 \citep{Kunitomo2021a}, which if varying can cause changes in the FUV luminosity. We use a power law disk model to compute disk parameters like the flaring angle. The gas surface density power law is given by,
\begin{equation}
    \Sigma_g(r) = \Sigma_0 \left(\frac{r}{r_0}\right)^{-p},
\end{equation}
where $\Sigma_0$ is the gas surface density at $r_0=1$ AU and the power law index $p=1$. We assume that the temperature is set by irradiation:
\begin{equation}
    T(r) = T_0 \left(\frac{r}{r_0}\right)^{-q},
\end{equation}
where $T_0$ is the temperature at $r_0 = 1$~AU and the power law index is $q=0.5$. We take $T_0 = 280$~K consistent with the Minimum Mass Solar Nebula model \citep{Hayashi1981}. The vertical profile of the gas follows a Gaussian distribution given by
\begin{equation}
\rho_g(z) = \frac{\Sigma_g}{\sqrt{2\pi}H}\mathrm{exp}\left(\frac{-z^2}{2H^2}\right),
\end{equation}
where $z$ is the vertical height above the midplane, and $H$ is the gas scale height given by $H=c_s/\Omega$, with $\Omega$ being the Keplerian frequency, and $c_s$ the speed of sound given by $c_s = (k_B T/m_g)^{1/2}$ where, $k_B$ is the Boltzmann constant and $m_g = 2.34m_H$ is the mean mass of a gas molecule. 
Table \ref{tab:diskparams} lists the disk parameters used in the fiducial model of the simulation. 
\begin{table}
    \centering
    \begin{tabular}{c c}
         \hline
         \hline
         Disk Parameters &  \\
         \hline
         $\alpha$-viscosity & $10^{-3}$ \\
         Temperature at 1 AU $T_0$ & 280 K \\
         dust to gas ratio & 0.01 \\
         Gas surface density at 1 AU $\Sigma_{0}$ & 1000 g/cm$^2$ \\
         FUV Luminosity $L_{\mathrm{FUV}}$ & 0.01 $L_\odot$ \\
         initial carbon fraction $f_c$ & 0.5 \\
         fragmentation velocity $v_{\mathrm{frag}}$ & 100 cm/s \\
         silicate material density $\rho_{\mathrm{sil}}$ & 3.3 g/cm$^3$\\
         carbonaceous material density $\rho_{\mathrm{c}}$ & 1.5 g/cm$^3$\\
         Surface density power law $p$ & 1 \\
         Temperature power law $q$ & 0.5 \\
         Pressure index $\eta$ & 0.01 \\
         \hline
         \hline
    \end{tabular}
    \caption{The parameters for the fiducial disk model}
    \label{tab:diskparams}
\end{table}
\subsection{Dust model}\label{sec-dustmodel}
A dust grain in our model is made up of silicate and carbonaceous material. We assume that a dust grain has a silicate core that is covered by carbonaceous material \citep{Draine2021,Kissel1987}. We take the densities of the silicate component (astronomical silicates) $\rho_{\mathrm{sil}} = 3.3$ g/cm$^3$ \citep{Draine2003} and the carbonaceous component (refractory organics) to be $\rho_c = 1.5$ g/cm$^3$ \citep{jager1998}. The total internal density of a dust particle is given by
\begin{equation}\label{eq:rhos}
    \rho_{\mathrm{s}} = \left(\frac{f_c}{\rho_c} + \frac{1-f_c}{\rho_{\mathrm{sil}}}\right)^{-1},
\end{equation}
where $f_c$ is the carbon fraction which is given by,
\begin{equation}
    f_c = \frac{m_c}{m_c + m_{\mathrm{sil}}},
\end{equation}
where $m_c$ is the mass of carbonaceous and $m_{\mathrm{sil}}$ is the mass of silicates in the dust grains.

We also track the residence time of the particle, which is the time spent by a particle in the exposed layer, i.e. the region above $\tau_z =1$ as defined by \citet{Klarmann2018}. It is very helpful to examine the regime of photolysis as discussed in Sec. \ref{sec:photolysis}. We calculate this in our simulation by tracking the total time spent by a particle at the exposed layer and the number of times a particle crosses the boundary to the exposed layer i.e.,
\begin{equation}\label{eq:restime}
    t_{\mathrm{res}} = \frac{t_{\mathrm{res,total}}}{n_\mathrm{res}}.
\end{equation}

\begin{figure}
    \centering
    \includegraphics[width= \columnwidth ]{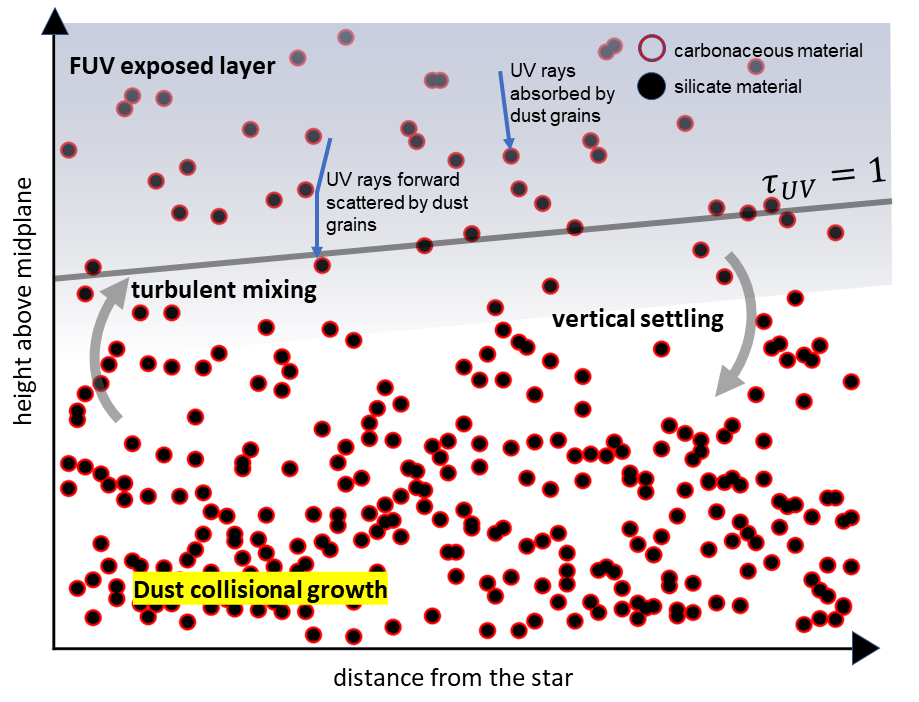}
    \caption[Vertical profile of the 1D disk model]{Figure shows the vertical slice of the model in use. At lower $z$ the larger dust aggregates settle down to the midplane. The dust particles move around due to the mixing processes. The blue gradient represents the FUV flux that gets absorbed and scattered as we come down to the midplane.}
    \label{fig:1D_model}
\end{figure}

Even though we do not set a hard boundary for our flux model as in \citet{Klarmann2018}, calculating the residence time $t_{\mathrm{tres}}$ is very helpful in order to place our work in the context of analytical models of carbon depletion where the depletion process is one of two regimes, residence time limited photolysis and unrestricted photolysis \citep{Klarmann2018}. Figure \ref{fig:1D_model} illustrates the vertical profile of our disk model, with the mixing processes: vertical settling and turbulent diffusion, formation of dust aggregates through collisional growth, and photolysis in the exposed layer.

\subsubsection{Dust distribution}
\begin{figure*}
    \centering
    \includegraphics[width = 0.8\textwidth]{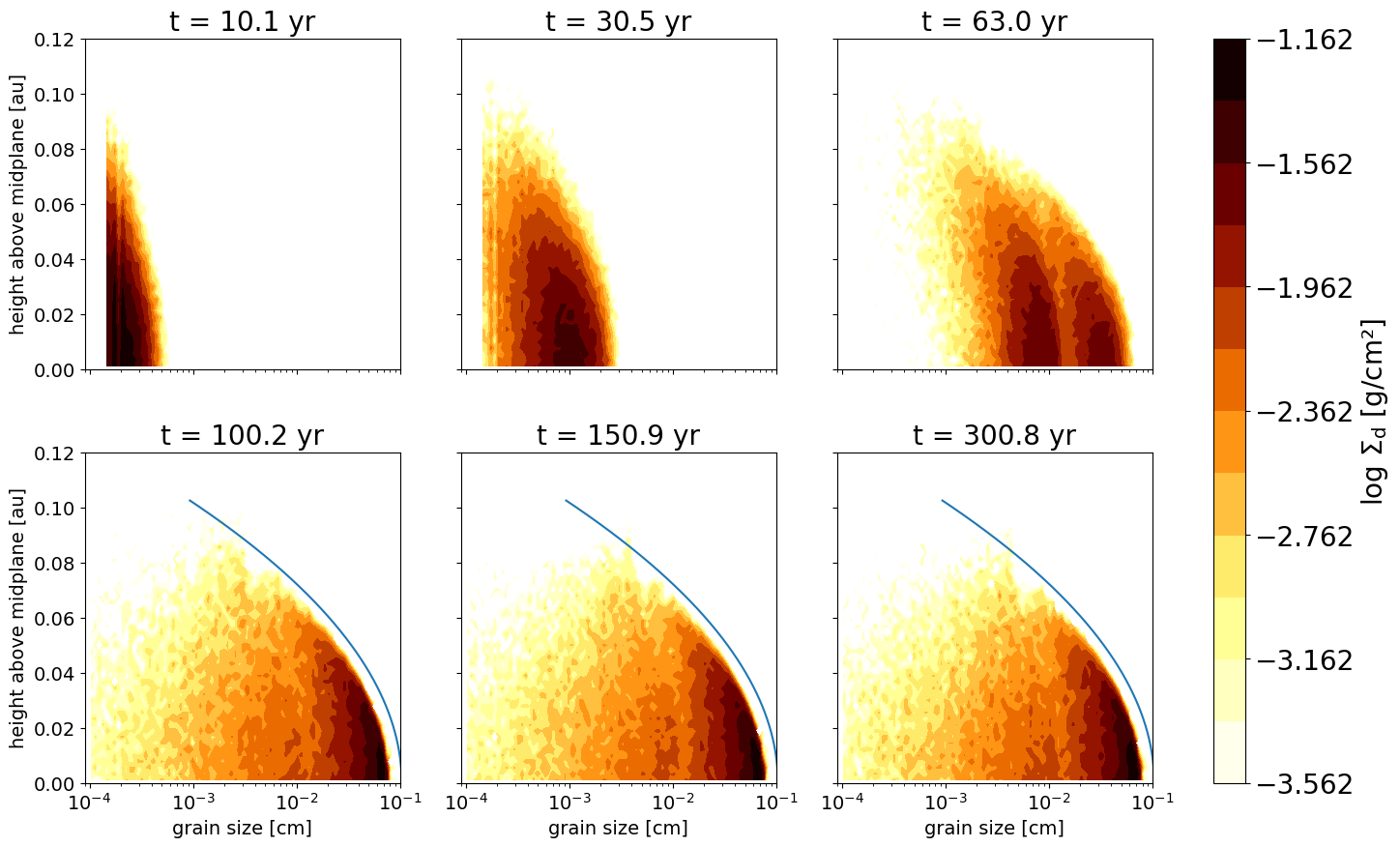}
    \caption[Dust surface density evolution]{Evolution of the dust surface density for a simulation with fragmentation and no bouncing. The blue lines denote the fragmentation barrier. }
    \label{fig:dustdens}
\end{figure*}

\begin{figure*}
    \centering
    \includegraphics[width = 0.8\textwidth]{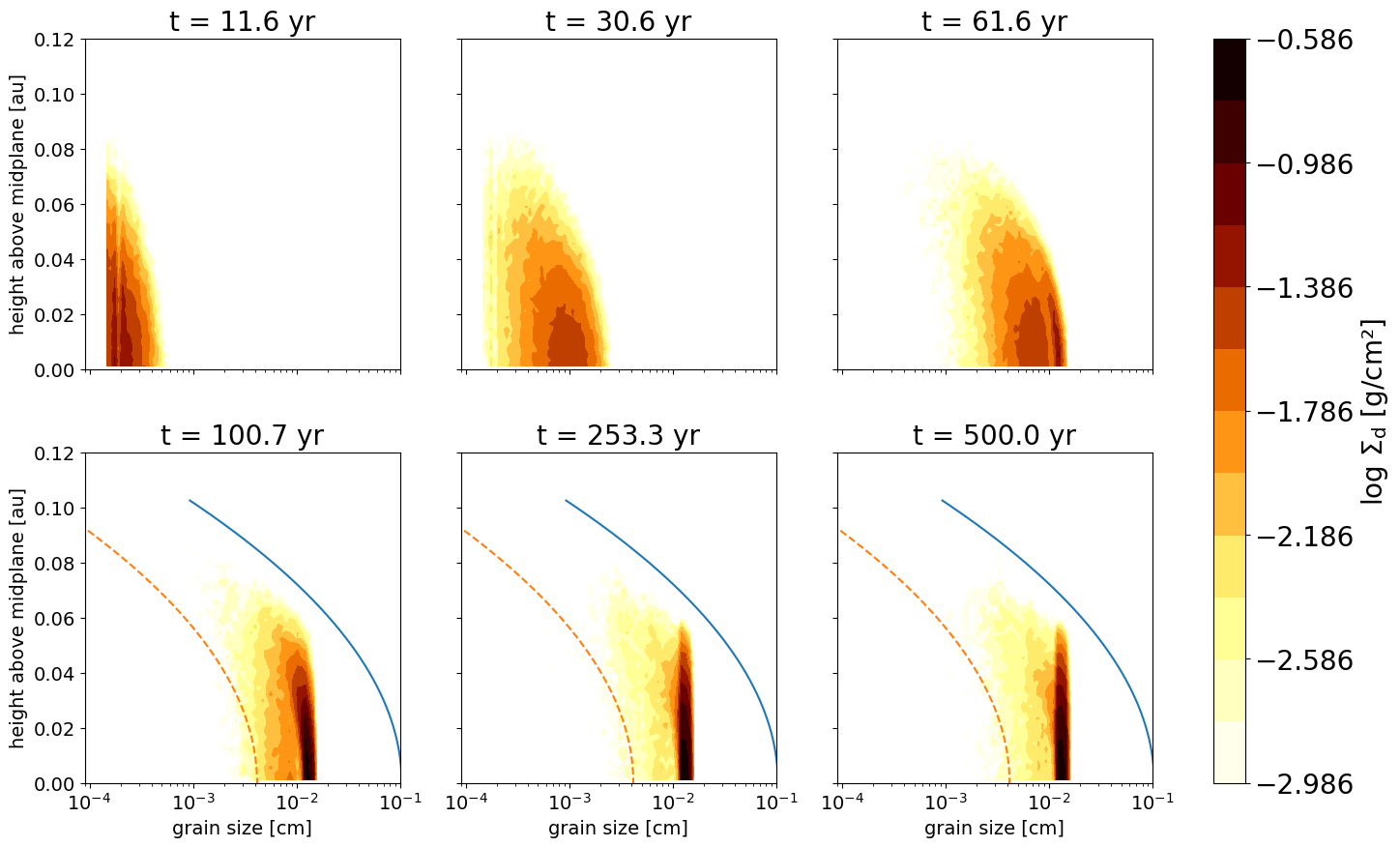}
    \caption[Dust density evolution with bouncing]{Evolution of the dust surface density when bouncing is included. The blue line is the fragmentation barrier and the dotted orange line is the bouncing barrier.}
    \label{fig:densbouncing}
\end{figure*}

We employ a simulated steady-state dust size distribution as an input to the photolysis runs. The initial dust distribution must be physical enough to switch on photolysis. Otherwise, we may have indeed ``solved'' the problem but for a non-physical distribution where all the grains have been very small. As discussed in Sect. \ref{sec:1DMC}, we implement three dust collisional outcomes: growth of dust particles by sticking, bouncing off each other, and fragmentation of dust particles. The latter two dust collisional outcomes limit the growth of the particles, meaning that they set barriers to the particle size. We performed two separate simulations: one with bouncing turned off and the other with bouncing turned on, to understand the effects of the different collision outcomes. Figure \ref{fig:dustdens} shows the time evolution of particle size distribution in a simulation without bouncing. We start with a monomer of size $a_0 = 1~\mu$m and let the setup evolve. We can see that after a few hundred years, the dust distribution settles into a coagulation-fragmentation equilibrium \citep{Birnstiel2011a}. The distribution follows the fragmentation barrier which limits the maximum grain size to \citep{Birnstiel2012}

\begin{equation}\label{eq:fragbarrier}
    a_{\mathrm{frag}} = f_\mathrm{f} \sqrt{\frac{8}{\pi}} \frac{\rho_g c_s}{ \rho_s \Omega_K}\mathrm{St}_{\mathrm{frag}},
\end{equation}

where $f_\mathrm{f}$ is a parameter of order of unity, and $\mathrm{St}_{\mathrm{frag}}$ is the Stokes Number at the fragmentation barrier which is given by,
\begin{equation} \label{eq:stfrag}
    \mathrm{St}_{\mathrm{frag}} = \frac{1}{3\alpha}\frac{v^2_{\mathrm{frag}}}{c_s^2}
\end{equation}
where $v_{\mathrm{frag}}$ is the fragmentation velocity. 
From Fig.~\ref{fig:dustdens}, we can see that the grains reach their maximum size $a_{\mathrm{frag}}$ at around 100~yrs. This is consistent with the coagulation timescale given by \citep{Birnstiel2012}
\begin{equation}
    t_{\mathrm{growth}}^{a_{\mathrm{frag}}} = t_{\mathrm{growth}} \mathrm{ln}\left(\frac{a_{\mathrm{frag}}}{a_0}\right) \sim 110\ \mathrm{yrs},
\end{equation}
where $t_{\mathrm{growth}}^{a_{\mathrm{frag}}}$ is the time taken by the dust to grow to size $a_{\mathrm{frag}}$ from monomer size $a_0$.
Similarly, for the bouncing barrier, we can modify Eq. \ref{eq:fragbarrier} as \citep{Dominik2024}

\begin{equation}\label{eq:bouncingbarrier}
    a_{\mathrm{bounce}} = f_\mathrm{f} \sqrt{\frac{8}{\pi}} \frac{\rho_g c_s}{ \rho_s \Omega_K}\mathrm{St}_{\mathrm{bounce}},
\end{equation}

where $f_\mathrm{f}$ is a parameter of order of unity, and $\mathrm{St}_{\mathrm{bounce}}$ is the Stokes Number at the bouncing barrier which is defined similarly to Eq.~\ref{eq:stfrag} where the fragmentation velocity $v_{\mathrm{frag}}$ is replaced with  $v_{\mathrm{bounce}}$, the bouncing threshold velocity. Typically, $v_{\mathrm{bounce}} < v_{\mathrm{frag}}$ means that a bouncing limited distribution leads to particles smaller than the fragmentation limited distribution. This can be indeed seen in Fig.~\ref{fig:densbouncing}. The distribution has a lower number of small and large particles. Bouncing reduces the probability of the grains fragmenting back to smaller sizes. The particles grow until the bouncing regime, once they reach there they do not grow more nor do they fragment (except for a small probability as mentioned in Sec. \ref{sec:1DMC}). This results in a distribution that is concentrated in the intermediate sizes unless processes like mixing advect large particles from the midplane to higher $z$ leading to fragmentation. In Fig.~\ref{fig:densbouncing}, this is one of the major sources for producing smaller dust grains and also the reason for the ``wall'' like structure. The largest grains from the midplane which are the largest particles of the column vertically mix and create a ``wall'' and once they reach the fragmentation barrier they fragment to produce smaller grains. This structure can be seen in the bottom row of Fig.~\ref{fig:densbouncing} with the dust surface density being concentrated in a short distribution of intermediate sizes across the vertical column until the fragmentation barrier.

\section{Results}\label{sec:results}
In this section, we will look at several aspects of the results that lead to carbon destruction in disks through photolysis. We also do a parameter study to look at how they affect the results. In Sect.~\ref{sec:discussion}, we interpret the physical implications of the results arrived at and place them in the context of planet formation.

\begin{table*}
    \centering
    \begin{tabular}{cccccccc}
    \hline
    \hline
         Simulation & $\alpha$ & mixing & collisions & $v_{\mathrm{frag}}$ & $f_c$ & bouncing & $v_{\mathrm{bounce}}$  \\
         \hline
         fiducial & $10^{-3}$ & yes & yes & 100 cm/s &  0.5 & no & N.A\\
         
         nomixfiducial & $10^{-3}$ & no & yes & 100 cm/s &  0.5 & no & N.A\\
         
         nocollfiducial & $10^{-3}$ & yes & no & 100 cm/s &  0.5 & no & N.A\\
         
         fiducialfc025 & $10^{-3}$ & yes & yes & 100 cm/s &  0.25 & no & N.A\\
    \hline
    \hline
    \end{tabular}
    \caption[Fiducial run parameters for looking at mixing and collision effects]{Parameters of the simulations that we use to discuss the fiducial run and the effects of collisions and mixing in the carbon depletion process}
    \label{tab:paramstudy1}
    
\end{table*}
\subsection{Results of the fiducial run}\label{sec:fiducialresults}
\begin{figure*} [ht]
    \centering
    \includegraphics[width = 0.9\textwidth]{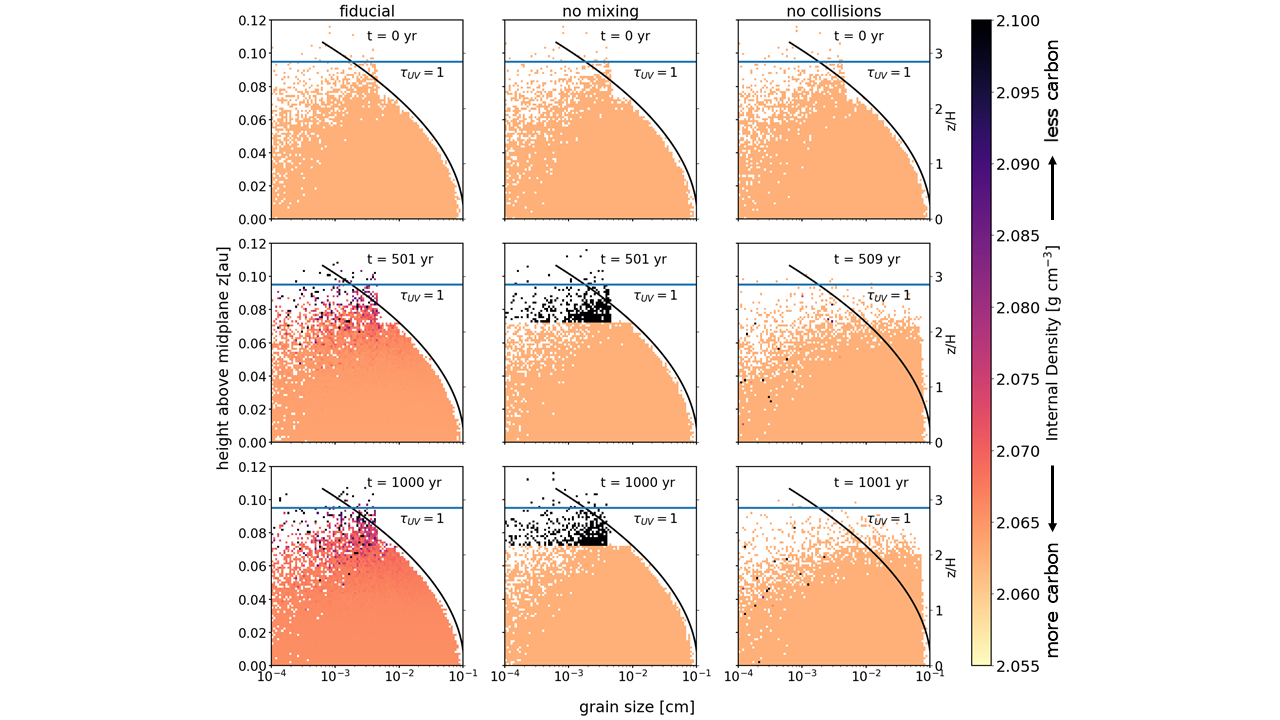}
    \caption[Mixing and collisional processes of the fiducial run]{(Left) Time evolution of the \texttt{fiducial} run (Centre) Time evolution of the \texttt{nomixfiducial} run where there is no vertical mixing (Right) Time evolution of the \texttt{nocollfiducial} run where there is no collisions.}
    \label{fig:photolysis2}
\end{figure*}

We now look at the results of the fiducial run whose disk parameters are given in Table \ref{tab:diskparams}. We take the carbon fraction $f_c = 0.5$ as mentioned in Table \ref{tab:paramstudy1} as a starting point to understand the working of photolysis. We make use of the analytical FUV model described in Sect.~\ref{Sec-FUV}. The left column in Fig.~\ref{fig:photolysis2} shows the time evolution of the simulation in terms of the internal density of dust aggregates which scales with the carbon fraction with higher densities corresponding to lower carbon fraction (see Eq.~\ref{eq:rhos}). As discussed in Sect.~\ref{Sec-simtest}, we evolve the dust to obtain a steady-state size distribution before turning on photolysis. It can be seen that the carbonaceous material in the grains higher up the disk midplane gets depleted immediately. This is a direct result of the fact that the FUV flux has larger magnitudes at lower optical depths. And as time evolves, we can observe that the carbon is depleted all through the vertical column albeit slowly through the homogeneous color change in the internal density of all the particles. Since the carbonaceous part of the grains is a coating to a silicate core as described in Sect. \ref{sec-dustmodel}, the carbonaceous coating can be shielded from the FUV photons due to sticking of the grains. We do not take the shielding into account as the particles in the exposed layer as seen in  Fig. \ref{fig:photolysis2} are very small. These particles are mostly monomers and small aggregates and the effect of such shielding in these particles will be very minimal. 

There are multiple processes involved in depleting and disseminating carbonaceous material. Photolysis by the FUV photons depletes carbon whereas processes like mixing and collisions redistribute the carbon locked in the grains close to the midplane. It is important to know how these processes affect the rate of photolysis. To explore the roles played by these processes we do a simulation that does not include vertical mixing ($\texttt{nomixfiducial}$ in Table \ref{tab:paramstudy1}) and a simulation that does not include collisional growth ($\texttt{nocollfiducial}$ in Table \ref{tab:paramstudy1}). Fig.~\ref{fig:photolysis2} shows the time evolution of the simulations for the fiducial run, without mixing and without collisions respectively. Both the no mixing case and the no collisions case are slower in carbon depletion. In the no-mixing case, we can see that only the upper layer is severely depleted in carbon compared to the midplane. The lack of mixing here fails to bring the carbon-rich grains closer to the midplane up to the exposed layer which greatly slows down the carbon depletion as shown in \citet{Klarmann2018}, but the collisions redistribute the carbon within the final grid and the grains get evenly depleted. This establishes the role of collisional redistribution of carbon. But one more thing to note from the no-mixing case is that how we bin particles might play a role in the results observed. Our model bins particles into grids of equal particle number and these grids are non-interacting. Having non-interacting grids might lead to unphysical results which bring out the need for a gridless tree code or a SPH-type code structure for the grids. We leave this for future work. In the no collisions case, where mixing is the only distribution mechanism for carbon, we can see that the grains indeed get mixed up to higher vertical heights, the grains that go up to larger heights are carbon-depleted grains that do not get replenished due to the lack of a carbon redistribution mechanism. This further stresses the role of collisional growth in accelerating carbon depletion. 

\begin{figure*}[ht]
    \centering
    \includegraphics[width = 0.8\textwidth]{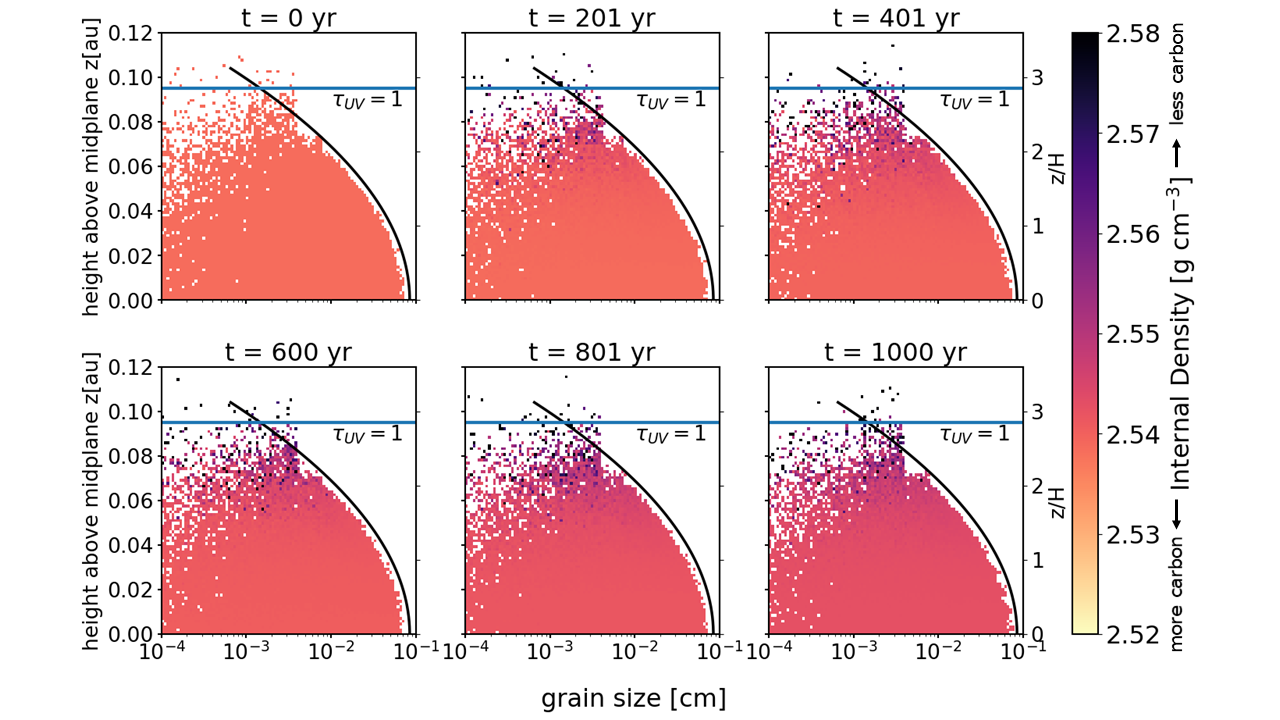}
    \caption[Time evolution of the fiducial run fc=0.25]{Time evolution of the \texttt{fiducialfc025} run with carbon fraction $f_c = 0.25$}
    \label{fig:DALInobouncealpha1e-3}
\end{figure*}

Now, we have established the working of the processes involved in the simulation. We now need to set the carbon fraction to physical values as in the early Solar System which is $f_c = 0.25$ following \citet{Klarmann2018} \citep[it is also close to the estimate $f_c=0.29$ by][]{Binkert2023}. Fig.~\ref{fig:DALInobouncealpha1e-3} shows the time evolution of the simulation $\texttt{fiducialfc025}$. 

In order to measure the carbon destruction timescale, \citet{Binkert2023} formulated the depletion timescale given by
\begin{equation}
    t_c  = \frac{\Sigma_c}{\Dot{\Sigma}_c},
\end{equation}
where $t_c$ is the carbon depletion timescale, $\Sigma_c$ is the surface density of the carbon grains and $\Dot{\Sigma}_c$ is the carbon decretion rate. This gives the timescale in which the carbon in the disk is completely depleted. It is a very helpful quantity in knowing how fast we are in depleting the carbon in the inner disk. This value is constant for an analytical estimate but for a simulation or a realistic case the same would not apply, so we estimate the carbon destruction timescale at every time snapshot $t_{\mathrm{snap}}$ as
\begin{equation}
    t_c(t) = - \frac{t_{\mathrm{snap}}}{\mathrm{log}\left(\frac{\Sigma_{c,t}}{\Sigma_{c,0}}\right)}.
\end{equation}

Numerically, to arrive at the depletion timescale we need to let the system settle into an "steady-state" that balances the collisional growth, mixing and photolysis. The value it settles down to is $t_{c,\mathrm{FUVmodel}} = 148.618 \pm 0.797$ kyr. This is the value arrived at after using the analytical FUV flux model. It means that the carbon reservoir is completely depleted by the time $t=t_c$. When the same calculation is arrived at with a simulation using the $\texttt{DALI}$ model as described in Sec. \ref{subsec-DALI} we arrive at a value, $t_{c,\mathrm{DALI}} = 116.610 \pm 0.764 $ kyr. This confirms that the DALI model indeed is more effective in depleting carbon as discussed in Sec. \ref{subsec-DALI}. The analytical model described in Sec. \ref{subsecFUV-analytical} is a lower limit of a flux model that includes scattering effects and a close estimate that helps with easy parameter exploration and saves computation time.

Similarly for the $\texttt{fiducial}$ run where $f_c = 0.5$, the carbon depletion timescale $t_{c,\mathrm{FUVmodel}} = 204.154 \pm 0.371$ kyr. This is a result that is different from \citet{Binkert2023} where their model gives us a constant carbon depletion timescale irrespective of the initial $f_c$ value. This difference arises due to the differences in the dust model. Our dust model as described in Sect.~\ref{Sec-simtest} assumes that all the carbon is present on the grains as a coating on the surface of the grains, which means every FUV photon interacts with a carbon atom. This automatically results in a faster $t_c$ for lower $f_c$. Whereas in \citet{Binkert2023}, they treat carbon and silicon as individual atoms which leads to FUV photons interacting with silicon atoms and as $f_c$ decreases the likelihood of the FUV photons to interact with the silicon atoms increases slowing down the otherwise what must be a quicker process as $f_c$ reduces thus keeping the carbon depletion timescale constant irrespective of the initial $f_c$ value.

\subsection{Regimes of photolysis}
In previous models described in section \ref{sec:photolysis}, there were two regimes of photolysis in disks defined in previous models \citep{Klarmann2018,Binkert2023}. The first one is the residence time-limited photolysis, where the carbon removal through photolysis is restricted by the residence times of the particles, that is the time grains spend in the exposed layer (see Fig.~\ref{fig:1D_model} for an illustration of the vertical profile with the exposed layer). Here the mixing processes are not efficient enough to replenish the carbon in the exposed layer due to the residence times being longer than the time it takes to fully deplete carbon. The second regime is unrestricted photolysis where the mixing is efficient enough to deliver fresh material to the exposed layer before all the carbon there is depleted. These were all applied to analytical models and as mentioned in \ref{sec:photolysis}. Here we calculate the quantities in our simulation to connect to the analytical model. We examine the regimes of photolysis in our model, which involves dust evolution along with a flux model that does not have a sharp boundary at $\tau_r=1$.

\subsubsection{Residence time of the particles}\label{sec-restime}
\begin{figure*}
    \centering
    \includegraphics[width = 0.9\textwidth]{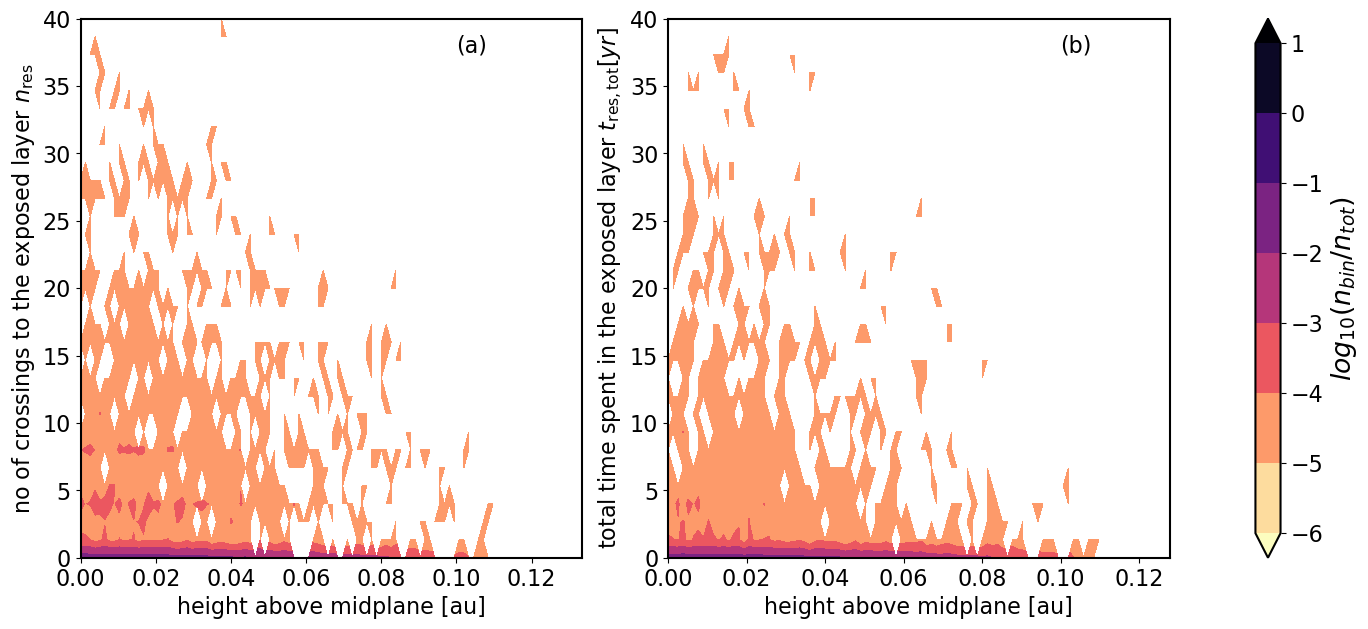}
    \caption[Total residence times and number of crossings to the exposed layer]{Panel (a): Number of particles (normalized with the total number of particles) at every height $z$ and their cumulative number of crossings to the exposed layer $n_{\mathrm{res}}$ at the end of the run \texttt{fiducialfc025}. Panel (b) Similar plot, but with total residence times of the particles $t_{\mathrm{res,tot}}$ at every height z.}
    \label{fig:restime}
\end{figure*}

\begin{figure}
    \centering
    \includegraphics[width = \columnwidth]{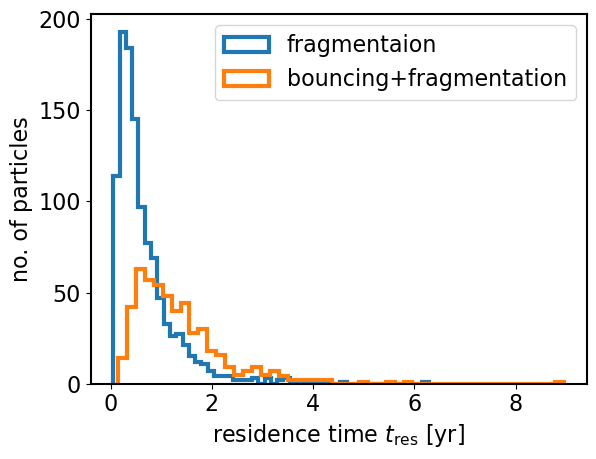}
    \caption[Histogram of the residence times of particles]{Histogram of the residence time $t_{\mathrm{res}}$ of the particles as calculated by Eq. \ref{eq:restime} for the runs \texttt{fiducialfc025} and \texttt{bouncefiducialfc025}.}
    \label{fig:tres}
\end{figure}

As mentioned in Eq. \ref{eq:restime}, we obtain the residence time of the particles from the total time a grain spends in the exposed layer $t_{\mathrm{res,total}}$  and the number of crossings to the exposed layer a particle does, $n_{\mathrm{res}}$. Fig.~\ref{fig:restime}a and Fig.~\ref{fig:restime}b show the number of crossings to the exposed layer and the total residence time versus the height above the midplane. It can be seen that the majority of grains settle to the midplane and never arrive at the exposed layer but some grains have a very high $n_{\mathrm{res}}$ and $t_{\mathrm{res,total}}$, which means they hover around the exposed layer rather than completing a full cycle. There is also no clear correlation between $t_{\mathrm{res,total}}$ and $n_{\mathrm{res}}$, i.e, a larger $n_{\mathrm{res}}$ doesn't mean that $t_{\mathrm{res,total}}$ is automatically higher. This tells us how starkly different the trajectories are for different particles making residence a tricky quantity to measure. To statistically arrive at the residence time, we remove the outliers mentioned before to get the residence time of a typical particle in the vertical column. Fig.~\ref{fig:tres} shows the histogram of the residence times of the particles. We get $t_{\mathrm{res}} = 0.658$ yr as the median value. To check the physical correctness of the value, we calculate the growth timescale of the particles at z = 3H and we arrive at $t_{\mathrm{growth,3H}} \sim 1$yr. Comparing our values to the growth timescale which is essentially the collision timescale tells us how long the particles take to collide, grow and settle. These values are comparable but due to the noisy nature of the residence time values more investigation to arrive at the values is required. However, the values we obtain indeed indicate that we are towards the unrestricted regime of photolysis since the residence times are very short which results in efficient mixing and therefore carbon gets efficiently replenished in the exposed layer.

\subsubsection{Carbon fraction across the vertical column}
\begin{figure*}[ht]
    \centering
    \includegraphics[width=0.8\textwidth]{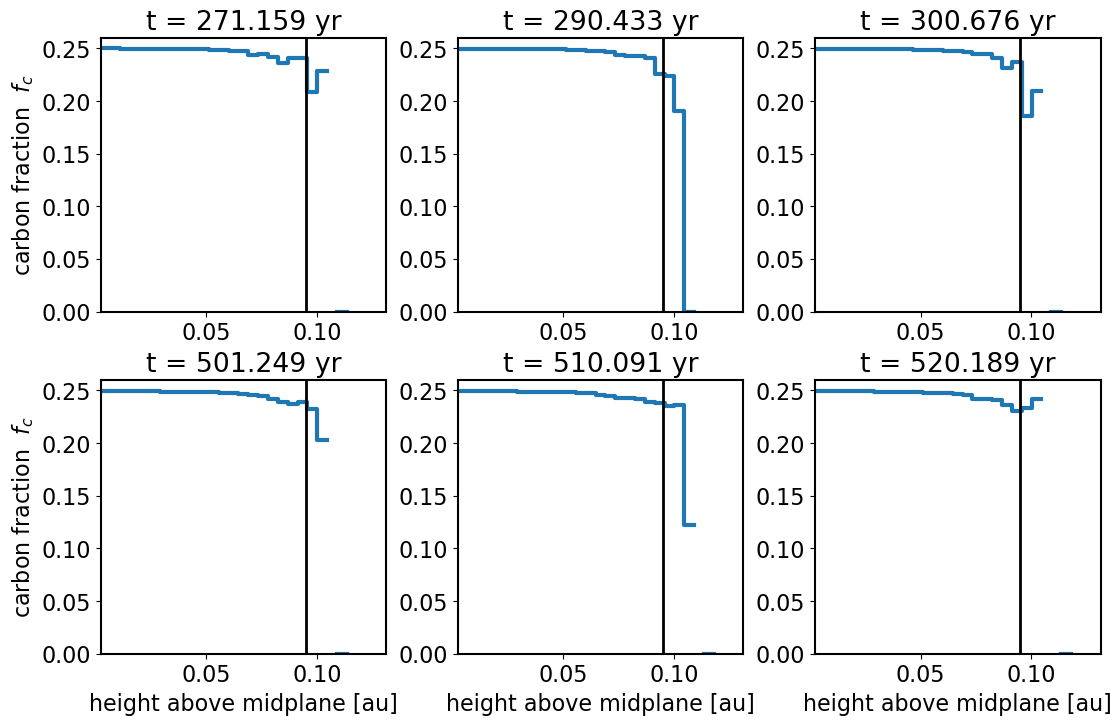}
    \caption[Carbon fraction at every height]{Top row and the bottom row show time evolution snaphosts of the carbon fraction $f_c$ for the height bins z for the run \texttt{fiducialfc025}. Moving from left to right we can see the carbon is depleted and replenished quickly. The vertical black line represents the $\tau_{\textrm{UV}} = 1$ line and the start of the exposed layer.}
    \label{fig:cfhist}
\end{figure*}

One more way to examine the photolysis regime is to look at the vertical profile of the carbon fraction of the particles. If we are indeed in the unrestricted regime, the carbon fraction should be distributed equally i.e. almost exactly as the initial conditions. But this will not be the case for residence time-limited photolysis since the grains that are depleted of carbon stay in the exposed layer creating a gradient in the vertical profile of the carbon fraction $f_c$. Fig.~\ref{fig:cfhist} shows the vertical profile of the carbon fraction $f_c$ mass averaged through the $z$ bins. At first glance the gradients in $f_c$ are visible, but if we look at the time slices close to the ones that have large gradients, mixing replenishes the carbon quickly. As an example in Fig.~\ref{fig:cfhist}, the bottom panel has three time slices where carbon fraction stays more or less the same at $f_c = 0.25$ in the first panel and gets a strong gradient as we move to the second panel and again comes back to an almost no gradient profile at $z = 0.10 AU$ in the final panel, i.e the depleted carbon is replenished due to mixing. In a fully residence time-limited regime the vertical gradient of the carbon fraction is very high, whereas in a completely unrestricted regime, the carbon fraction is uniform vertically. This points out that we are not in the residence time-limited regime, but it is still not strictly in the unrestricted regime due to the slight variations in the vertical profile.

Looking at the residence times and the vertical profile of the carbon fraction, it is evident that the simulation lies between a residence time-limited regime and an unrestricted regime. This is because the simulation has strong enough mixing (due to vertical mixing and collisional redistribution of carbon) to break the residence time barrier but not strong enough to become completely unrestricted. This can be considered a natural outcome of realistic setups where the interplay of multiple processes can create a mix of the two regimes.

\subsection{Introducing bouncing}\label{sec-bouncingphot}
 
\begin{figure*}
    \centering
    \includegraphics[width=0.8\textwidth]{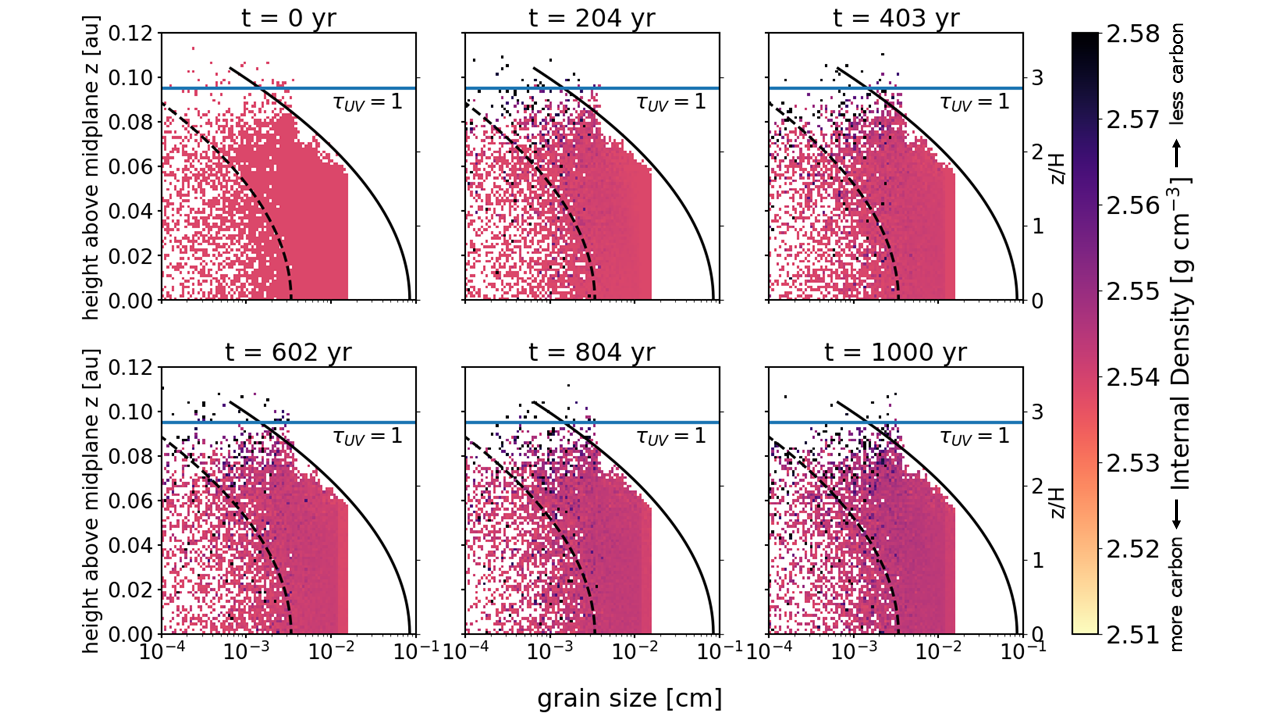}
    \caption[Fiducial run with bouncing included]{Time evolution of the $\texttt{bouncefiducialfc025}$ run. The distribution has a smaller maximum size than the fragmentation-limited distribution when it is bouncing-limited.}
    \label{fig:DALIvbounce20alpha1e-3}
 \end{figure*}

 So far, we analyzed the photolysis runs for models that only include growth by sticking and fragmentation as the only collisional barrier. However, experiments have shown that grains also bounce of each other. \citep{Guttler2010,Zsom2010}. As explained in Sect. \ref{sec-dustmodel}, we introduced bouncing as a collisional outcome in some of the models. Here, we used $v_{\mathrm{bounce}} = 20$ cm/s consistent with the results presented by \citet{Guttler2010}. Fig.~\ref{fig:DALIvbounce20alpha1e-3} shows the time evolution of the $\texttt{bouncefiducialfc025}$ run that introduces the bouncing barrier and has $\alpha = 0.001$. The initial size distribution, as motivated in Sect.~\ref{sec-dustmodel}, has a higher number of intermediate-sized grains. The reduction in the small grain population can affect the carbon depletion rate. If there are not enough small grains to efficiently mix the carbon up to the regions with enough FUV flux, the overall carbon depletion rate will be slowed down. This is evident when looking at the depletion timescale for the \texttt{bouncefiducialfc025}, which is $\tau_{\text{c,DALI}} = 162.285 \pm 0.546 \text{kyr}$ which is longer than the depletion times in the case of only fragmentation. Interestingly, when looking at the residence times of the particles, we arrive at a median value of $t_{\textrm{res}}=1.34 \;\textrm{yr}$. This is higher than the residence time value we get for the $\texttt{fiducialfc025}$ as seen in Sect.~\ref{sec-restime}. However, the number of particles that cross $\tau_{\textrm{UV}}=1$ line is lower than what we see in $\texttt{fiducialfc025}$, which is due to the reduced small grain population in the bouncing case. This means that a particle spends more time in the exposed layer before growing and settling into the midplane thereby increasing the residence times of the particles. In this case, the particle will be influenced by the strength of vertical mixing more than the growth to settle into the midplane.

\subsection{$\alpha$ parameter study}\label{sec-alphaparamstudy}
Our simulations use certain parameters that influence the physical processes happening in the disk. The turbulent viscosity parameter $\alpha$, which governs the strength of the mixing, is one of the most important ones. Constraining realistic values of $\alpha$-viscosity for disks from observations has always been a puzzle. \citet{Andrews2009} constrained $\alpha \approx 0.0005 - 0.08$ for the Ophiuchus star forming region. Studies of the TW Hya disk constrained the $\alpha$ values to $\alpha \sim 0.09$ as an upper limit \citep{Teague2016} and as a follow up \citet{Flaherty2018} arrived at $\alpha < 0.007$. More recent studies show that the turbulence levels vary from disk to disk. \citep{Paneque2024} constrained the turbulence level in IM Lup to be very high $\alpha \sim 10^{-1}$ and \citep{Villenave2022} observed a settled disk around Oph163131 with $\alpha \gtrsim 10^{-5}$. Proper constraints on getting the $\alpha$ viscosity value are still an ongoing discussion with a lot of model dependence and varying turbulence levels in disks. We explored the parameter space of $\alpha$ and considered values of 0.0001, 0.001, and 0.01. In order to observe the effects of $\alpha$ only on the strength of mixing and not on the collisional growth, we modified the fragmentation velocity to keep the maximum grain size constant following Eq. \ref{eq:fragbarrier} as
\begin{equation}\label{eq:vfragafrag}
    a_{\mathrm{frag}} \sim \frac{v_{\mathrm{frag}}^2}{\alpha}.
\end{equation}
\begin{table*}[ht]
    \centering
    \begin{tabular}{cccccccc}
    \hline
    \hline
         Simulation & $\alpha$ & mixing & collisions & $v_{\mathrm{frag}}$ & $f_c$ & bouncing & $v_{\mathrm{bounce}}$  \\
         \hline
         fiducialfc025 & $10^{-3}$ & yes & yes & 100 cm/s &  0.25 & no & N.A\\
         alpha1e-4fc025 & $10^{-4}$ & yes & yes & 31.6 cm/s &  0.25 & no & N.A\\
         alpha0.01fc025 & $10^{-2}$ & yes & yes & 316 cm/s &  0.25 & no & N.A\\
         bouncefiducialfc025 & $10^{-3}$ & yes & yes & 100 cm/s &  0.25 & yes & 20 cm/s\\
         bouncealpha1e-4 & $10^{-4}$ & yes & yes & 31.6 cm/s & 0.25 & yes & 9.45 cm/s \\
         bouncealpha1e-2 & $10^{-2}$ & yes & yes & 316 cm/s & 0.25 & yes & 42.2 cm/s\\
    \hline
    \hline
    \end{tabular}
    \caption[Parameters of the simulations done for the $\alpha$ parameter study]{Parameters of the simulations done for the $\alpha$ parameter study}
    \label{tab:paramstudy}
    \end{table*}
    
This way, limiting the fragmentation velocity ensures the dust distribution remains similar and mixing is the only process affected by modifying the $\alpha$ value. Similarly, the bouncing threshold velocities are modified as in Eq.\ref{eq:vfragafrag} to suit the alpha values, with the starting point being $v_{\mathrm{bounce}} = 20$~cm/s for $\alpha = 0.001$. Table \ref{tab:paramstudy} shows the simulations and the parameters we use for this parameter study. The fiducial runs have $\alpha = 0.001$ and now we look at runs with $\alpha = 0.0001$ \& $\alpha=0.01$.

\begin{figure*}
    \centering
    \includegraphics[width=0.8\textwidth]{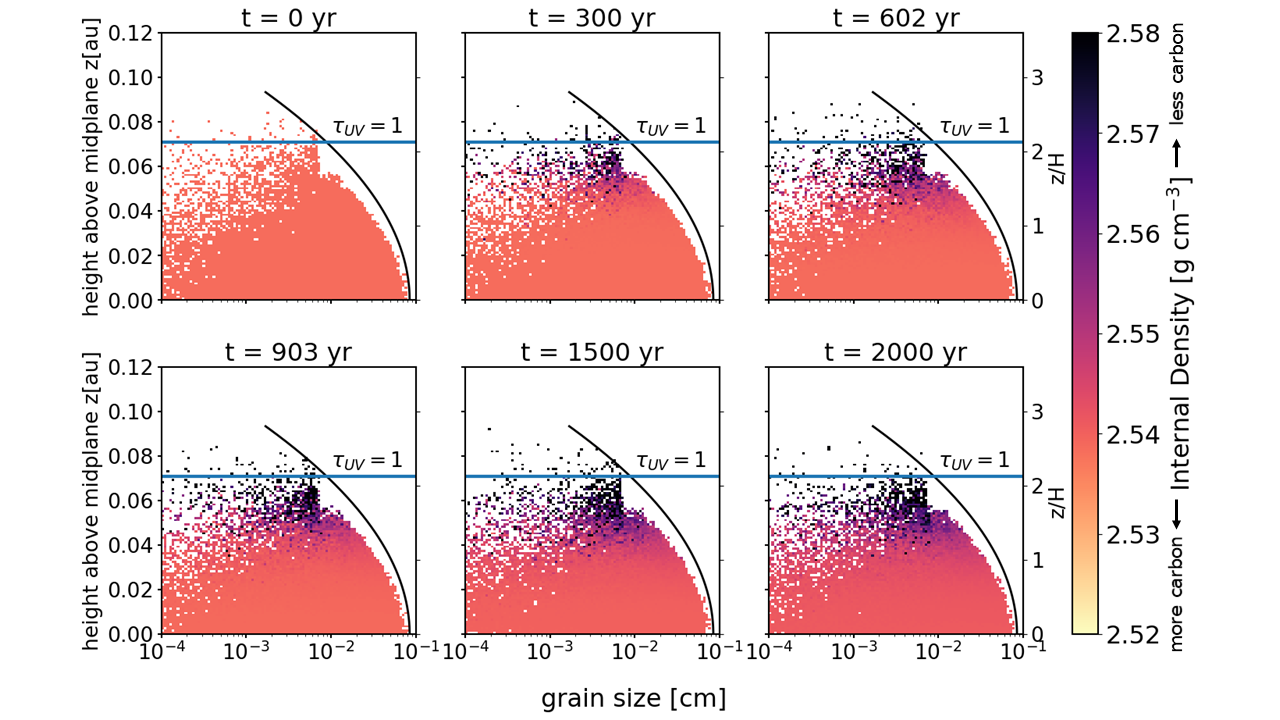}
    \caption[DALI simulation $\alpha = 0.0001$]{Time evolution of the $\texttt{alpha1e-4fc025}$ run where $\alpha = 0.0001$. We use the FUV flux from the DALI data to run the simulation}
    \label{fig:DALInobouncealpha1e-4}
\end{figure*}

\begin{figure*}
    \centering
    \includegraphics[width=0.8\textwidth]{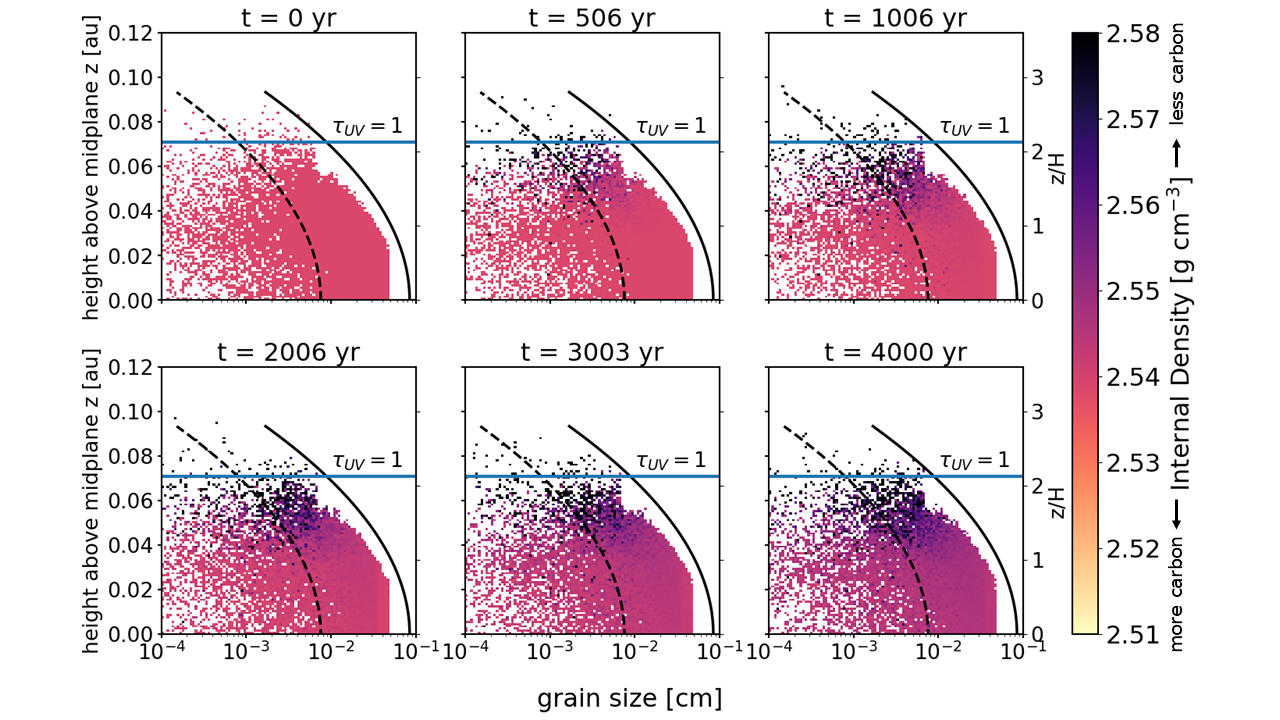}
    \caption[DALI simulation $\alpha=0.0001$ with bouncing]{Time evolution of the $\texttt{bouncealpha1e-4}$ run where $\alpha = 0.0001$ and we introduce the bouncing barrier. We use the FUV flux from the DALI data to run the simulation}
    \label{fig:DALIvbounce20alpha1e-4}
\end{figure*}

\begin{figure*}
    \centering
    \includegraphics[width=0.8\textwidth]{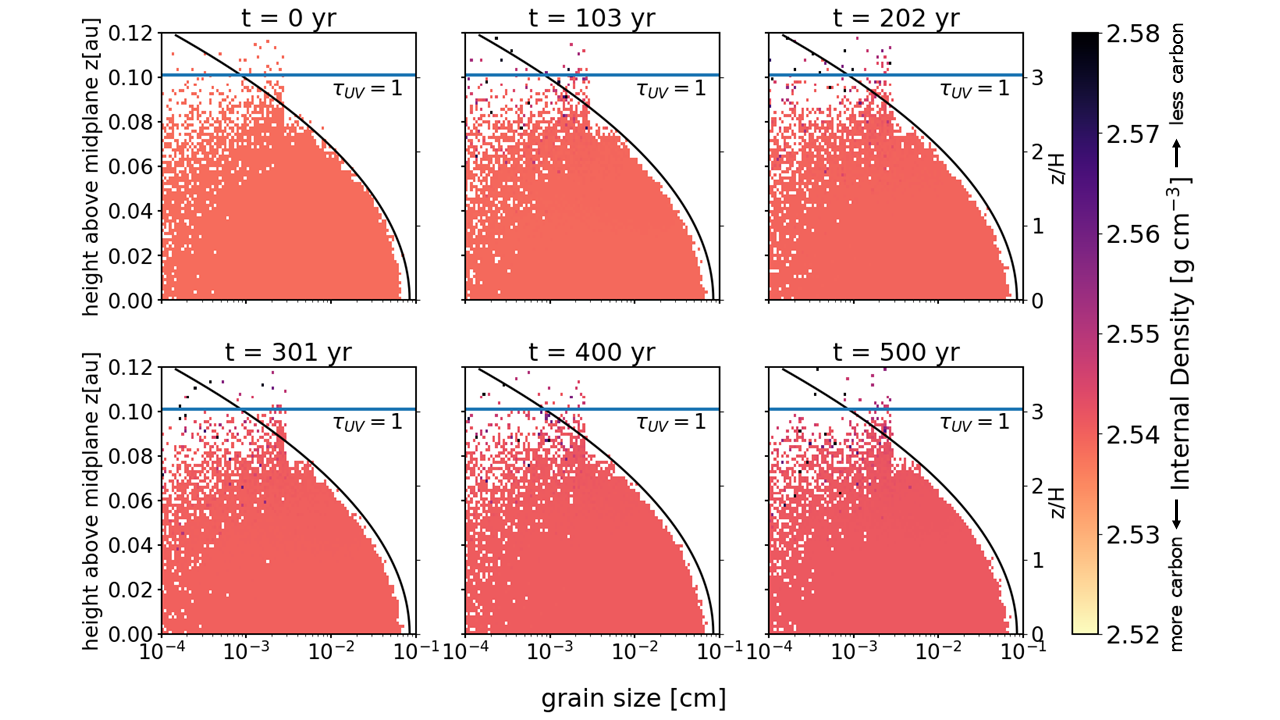}
    \caption[DALI simulation $\alpha = 0.01$]{Time evolution of the $\texttt{alpha0.01fc025}$ run where $\alpha = 0.01$. We use the FUV flux from the DALI data to run the simulation}
    \label{fig:DALInobouncealpha1e-2}
\end{figure*}
\begin{figure*}
    \centering
    \includegraphics[width=0.8\textwidth]{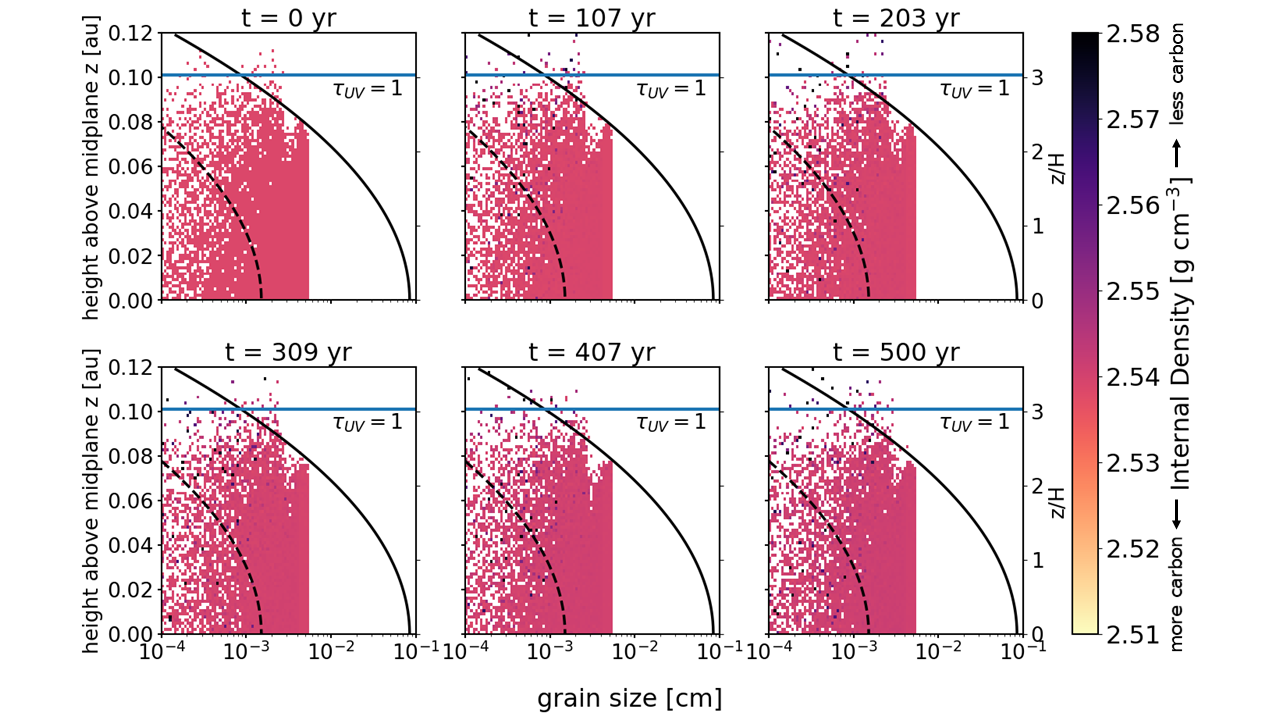}
    \caption[DALI simulation $\alpha = 0.01$ with bouncing]{Time evolution of the $\texttt{bouncealpha1e-2}$ run where $\alpha = 0.01$ and we introduce bouncing. The ``wall'' that arises due to the combination of mixing and the bouncing barrier is prominent with strong turbuence. We use the FUV flux from the DALI data to run the simulation}
    \label{fig:DALIvbounce20alpha1e-2}
\end{figure*}

Fig.~\ref{fig:DALInobouncealpha1e-4} shows the time evolution of the $\texttt{alpha1e-4fc025}$ run, where $\alpha = 0.0001$ and bouncing is not included. It can be seen that although the size distribution is the same, due to the weaker mixing strength the vertical column is more settled than in the case where $\alpha = 0.001$, which leads to a slower depletion of carbon. When bouncing is introduced for the $\alpha = 0.0001$ case, (refer $\texttt{bouncealpha1e-4}$ in Table \ref{tab:paramstudy}), we can see in Fig.~\ref{fig:DALIvbounce20alpha1e-4} that the distribution has a weaker ``wall'' when compared to the $\alpha=0.001$ case. Since the mixing is weaker here, the particles do not get mixed up until the fragmentation barrier. Fig.~\ref{fig:paramstudyplot}a shows the depletion of $f_c$ with time compared to the $\alpha = 0.001$ case with and without bouncing. Carbon loss is slower when there is weaker mixing. We now look at the case $\alpha = 0.01$ ($\texttt{alpha1e-2fc025}$ in Table \ref{tab:paramstudy}). In Fig.~\ref{fig:DALInobouncealpha1e-2}, we can see the time evolution of the run $\texttt{alpha1e-2fc025}$, where bouncing is not included. Here the mixing is stronger and the particles are mixed up to a higher vertical height compared to the models with weaker turbulence. This leads to a faster depletion of carbon. Similarly, when bouncing is included, due to stronger mixing, we get the most prominent ``wall'', i.e.~the mixing of the grains leading to crossing the fragmentation barrier is much more effective here than in the weaker turbulence runs. This stresses the fact that when including bouncing and fragmentation, the strength of mixing plays a very important role in determining the dust distribution. Fig.~\ref{fig:paramstudyplot}a shows the depletion of $f_c$ with time compared to all the other turbulence cases. Out of all the runs, $\alpha$ = 0.01 leads to the fastest carbon depletion. When looking at the residence times of particles for the runs with different $\alpha$, for $\alpha=0.01$ we get residence times of $t_{\textrm{res}}=0.13\;\textrm{yr}$ for the run $\texttt{alpha1e-2fc025}$ where bouncing is not included and $t_{\textrm{res}}=0.26\;\textrm{yr}$ for the run $\texttt{bouncealpha1e-2}$ reaffirming the increased residence times for particles in the bouncing regime as seen in Sect.~\ref{sec-bouncingphot}. And for $\alpha=0.0001$ we see very few particles (~10) crossing the exposed layer thus pointing to a very settled distribution and weak mixing for both \texttt{alpha1e-4fc025} where bouncing is not included and in \texttt{bouncealpha1e-4} where bouncing is included.

\section{Discussion}\label{sec:discussion}
As we describe in Sec. \ref{sec:results}, the outcome of dust evolution and carbon depletion by photolysis change depending on the different parameters within our model. It is now time that we put that in the context of carbon depletion timescales and look at the planet-forming conditions. To compare with analytical estimates, \citet{Binkert2023} arrived at 92.5 kyr as the depletion timescale for the unrestricted photolysis case and 4.2 Myr for the residence time-limited photolysis case. We estimate values of 116.610 kyr in Sect.~\ref{sec:fiducialresults}, which points towards the unrestricted photolysis in our fiducial run.

\begin{figure*}
    \centering
    \includegraphics[width=0.9\textwidth]{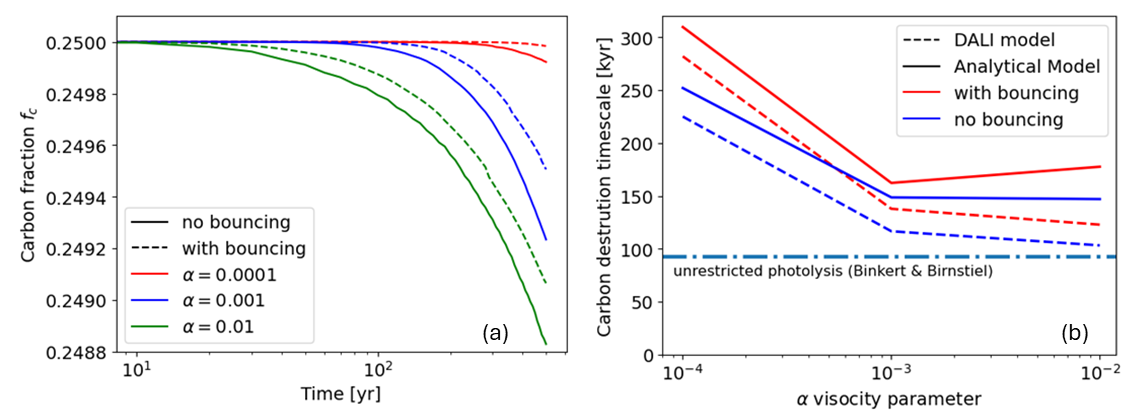}
    \caption[Carbon fraction and depletion timescales for parameter study]{Panel (a): Time evolution of the carbon fraction $f_c$ for the various parameters. Shown, are the runs done with the DALI FUV data. Panel (b): Carbon depletion timescales $t_c$ for the various values of turbulence strength $\alpha$. }
    \label{fig:paramstudyplot}
\end{figure*}

We calculated the carbon depletion timescales for the different simulations done in the $\alpha$ parameter study in Sec. \ref{sec-alphaparamstudy} as shown in Fig.~\ref{fig:paramstudyplot}a. In Fig.~\ref{fig:paramstudyplot}b we also show the timescales for runs with both the FUV models, that is the FUV flux from the DALI setup as described in Sect.~\ref{subsec-DALI} and the FUV analytical model from Sect.~\ref{subsecFUV-analytical}. While using the DALI model, there is a clear decrease in the depletion timescale with the increase in turbulence, which is not exactly the case for the analytical FUV model. Further studies into the effects of mixing for such a model are required. In general, the DALI model is faster in depleting the carbon in the disk. In the case of the residence time-limited photolysis, the flux model would have not mattered as the carbon was not replenished in the upper layers and the photons did not hit the carbon grains as they would in a replenished disk. But here it is clear that the carbon gets replenished quickly and depletion happens based on the FUV photons present in the model.

Another piece of evidence that points to an unrestricted photolysis model is that there is not much difference in the depletion time scale for the cases of $\alpha =0.001$ and $\alpha = 0.01$. This agrees with the definition of the unrestricted case where the strength of the mixing doesn't matter after mixing has reached the required efficiency. This marks the importance of collisional mixing as a means to redistribute carbon since we reach the spectrum of the unrestricted regime with a moderate value of $\alpha = 0.001$.

Considering all our $\alpha$ runs with the DALI model, we arrive at a destruction timescale range $\sim$ 100-280 kyr. As discussed in Sect.~\ref{sec:solarsystemcarb}, the meteorites in the inner Solar System are severely depleted in carbon pointing to the possibility that the inner Solar System was completely devoid of carbon before 1 Myr and the dust incoming from the outer disks brought the carbon with it later. The meteorites that formed early are even more depleted in carbon than later ones, necessitating a quick process to destroy carbon \citep{Gail2017}. Although the carbon destruction timescales derived in our models are quick enough to destroy carbon in a few hundred kyr, we need to take into account the particles that drift and replenish carbon into the disks. A useful quantity to measure the importance of the particles drifting is the radial drift timescale which is the time taken for a particle to reach the central star \citep{Birnstiel2012}. Even though the particle growth timescale is shorter than the radial drift timescale at 1~AU, our carbon depletion timescale is not lower than the radial drift timescale. This means that the carbon-rich particles from the outer disk will drift towards the inner disk and may replenish carbon faster than it is destroyed by photolysis. But recent studies point to a case where the planetesimals formed just beyond the water snow line for the early Solar System, binding the particles in the outer disk and leading to a bifurcation of planet formation histories of the inner and outer disk \citep{Lichtenberg2021}. This would lead to a weak radial mixing in the early Solar System.

We computed the time our models listed in Table~\ref{tab:paramstudy} take to reach the carbon fraction of the bulk earth as discussed Sec.~\ref{sec:solarsystemcarb}. This is computed by
\begin{equation}
    t_{f_c=0.01} =\frac{\Sigma_{c,f_c=0.01}- \Sigma_{c,0}}{\Dot{\Sigma_c}}.
\end{equation}
We obtained timescales ranging from 100 to 276~kyr. 

If the early planetesimals already formed before 300 kyr, our timescales derived might not have been fast enough. A process that destroys carbon faster is required to operate alongside photolysis. \citet{Li2021} discussed the sublimation of carbon in a hot inner disk which would be fast enough to completely deplete the carbon. They also look at outbursts like the FU Orionis \citep{Hartmann2003,Zhu2009}. These short outbursts lasting $\sim$100~years lead to viscously heated disks with higher temperatures in the disk midplane leading to sublimation of carbon. Such outbursts along with photolysis would be the likely combination to deplete carbon in the young Solar System and arrive at the present values. \citet{Binkert2023} discussed the effects of photolysis and sublimation together in a disk and adding sublimation (pyrolysis) decreases the depletion timescale by a factor of ten. Just extrapolating this from the timescales of 100-276 kyr leads to depletion timescales of $\sim$ 30 kyr, and this combination may be the key to solving the carbon depletion puzzle. Numerically modeling a viscous disk and outbursts along with sublimation would give us the complete and final picture of refractory carbon depletion.

One more important aspect that needs to be explored is the effects of dust grain substructure on the chemistry in disks. We have not implemented porous dust grains \citep{Ormel2007} and the various dust evolution outcomes related to porosity, for example, bouncing with compaction and mass transfer \citep{Guttler2010}. Exploring the dust properties and how they affect surface-level chemical reactions would give us more insight into the interplay of dust evolution and disk chemistry setting the stage for planet formation.

\section{Summary and conclusions}
Refractory material composition in the Solar System bodies gives a lot of clues about planet formation and planet-forming conditions \citep{Bosman2021}. Carbon in its refractory form is significantly depleted in the inner Solar System and the reason for this is unknown. We look at one of the solutions for the puzzle, carbon depletion by the photo-induced depletion process, photolysis. To examine the process, we used a one-dimensional Monte Carlo dust evolution code to simulate the dust evolution processes, and we introduced a simple analytical FUV model to estimate the FUV flux in disks. We also benchmarked the model with results from the Monte Carlo radiative transfer code DALI. The major results are summarized here as follows:
\begin{itemize}
    \item The analytical FUV model developed in Sect.~\ref{subsecFUV-analytical} agrees well with the results from the DALI code (Sect.~\ref{subsec-DALI}) for the optical depths where most of the carbon depletion occurs. This helps in creating a model to calculate the lower bounds for FUV fluxes including the scattering effects.
    \item When bouncing is introduced as a collisional outcome, the dust distribution is very different from the coagulation-fragmentation equilibrium. The grains do not grow as big as in the fragmentation-limited distribution. Due to bouncing, the grains do not fragment often which leads to a smaller population of small grains, which are mixed up to the upper layers where the photolysis is effective.
    \item To make photolysis an effective carbon depletion process, mixing and transfer of material within the disk is very important. We find that, along with the vertical mixing, collisional growth is also a very important process to redistribute carbon across the grain sizes and replenish carbon-rich material in the upper disk. 
    \item The photolysis process in our simulations is not residence time-limited, but it is not exactly unrestricted either. This necessitates the need to think about the photolysis regimes as a spectrum. Our fiducial simulation falls closer to the unrestricted regime.
    \item Our results show that with the increase in turbulence strength, the depletion timescales reach closer to the unrestricted regime. The depletion timescales range from 100-300 kyr with the higher values corresponding to lower $\alpha$.
    \item Our models suggest that photolysis would reduce the carbon fraction to the bulk Earth value of $f_c = 0.01$ within 100-276 kyr. This is faster than the previous models suggested. However, we do not take into account the radial mixing that might replenish carbon at the Earth location (although the isotopic dichotomy suggests that the early Solar System had weak radial mixing, see, e.g.~\citealt{Kleine2020}).    
\end{itemize}

Overall, our results suggest that carbon depletion by photolysis might have played an important role in shaping the dust evolution in the early Solar System but it is unlikely to have led to the low carbon fractions seen in the earliest formed meteorites and the Earth on its own.

\begin{acknowledgements}
V.V. and J.D. were funded by the European Union under the European Union’s Horizon Europe Research \& Innovation Programme 101040037 (PLANETOIDS). T.B. acknowledges funding from the European Union under the European Union's Horizon Europe Research and Innovation Programme 101124282 (EARLYBIRD) and funding by the Deutsche Forschungsgemeinschaft (DFG, German Research Foundation) under grant 325594231, and Germany's Excellence Strategy - EXC-2094 - 390783311. Views and opinions expressed are, however those of the author(s) only and do not necessarily reflect those of the European Union or the European Research Council. Neither the European Union nor the granting authority can be held responsible for them.
\end{acknowledgements}

\bibliographystyle{aa}
\bibliography{library}
\begin{appendix}
\section{FUV flux model}\label{subsecFUV-analytical}
\subsection{Ly$\alpha$ Scattering} \label{subsec:lya}
Ly-$\alpha$ flux is $\approx$ 80\% of the total stellar FUV flux which makes it important to look at the propagation of the Ly-$\alpha$ photons. \citet{Bethell2011} showed the scattering and propagation of Lyman-$\alpha$ photons in the context of protoplanetary disks. We make use of their model and construct a simple analytical prescription for the FUV flux. The Ly-$\alpha$ photons irradiated by the star get intercepted by the Hydrogen Layer (H-Layer) high up the disk and get isotropically scattered. This means that 50 \% of the Ly-$\alpha$ photons or 50 \% of the total Ly-$\alpha$ flux intercepted is scattered downward at the scattering surface which is denoted with the optical depth $\tau_r = 1$. Then, the Ly-$\alpha$ flux that is scattered downward to the vertical column at a height z and at a radius r of the disk is given by,
\begin{equation}\label{eq:lya}
    F'_{\mathrm{Ly\alpha}}(r,z) = 0.5 \frac{1}{e}F_{\mathrm{Ly\alpha}}(r) e^{-\tau_z(r)},
\end{equation}
where $F_{\mathrm{Ly\alpha}} = 0.8F_{\odot,\mathrm{FUV}}$ is the unimpeded Ly-$\alpha$ flux, the pre-factor $\frac{1}{e}$ accounts for the flux that is lost before reaching the scattering surface $\tau_r = 1$. The vertical optical depth $\tau_z$ is computed by,
\begin{equation}\label{eq:lytau}
    \tau_z = - \int^z_\infty \kappa_{\mathrm{Ly}\alpha} \rho_H(z') dz',
\end{equation}
where $\rho_H$ is the vertical density distribution of Hydrogen and $\kappa_0$ the opacity is given by,
\begin{equation}
    \kappa_{\mathrm{Ly}\alpha} = \frac{\sigma_{\mathrm{Ly}\alpha}}{m_H},
\end{equation}
where $m_H$ is the mass of a hydrogen atom and  $\sigma$ is the scattering cross-section for the Ly$\alpha$ photons by H atoms. We take $\sigma_{\mathrm{Ly}\alpha} = 10^{-21} \mathrm{cm}^{-2}$, following \citet{Bethell2011}. To calculate the location of the scattering surface $\tau_r = 1$ we apply
\begin{equation} \label{eq:taur}
    \tau_r \approx \frac{1}{\Phi} \tau_z,
\end{equation}
where $\tau_z$ is the vertical optical depth and $\Phi$ is the flaring angle of the disk which is given by \citep{Chiang1997,Chiang2001}
\begin{equation} \label{eq:flaring}
    \Phi = 0.4\frac{R_*}{r} + r\frac{d}{dr}\left(\frac{H_s}{r}\right),
\end{equation}
where $H_s = 4h$ and $h = \frac{c_s}{\Omega}$ is the scale height.

This way of calculating the radial optical depth $\tau_r$ is verified following the method of \citet{Watanabe2008}, details of which can be found in Appendix \ref{sec:radoptdepth}. 

\subsection{FUV scattering} \label{subsec:fuvscat}

\begin{figure}
     \centering
     \includegraphics[scale =0.35]{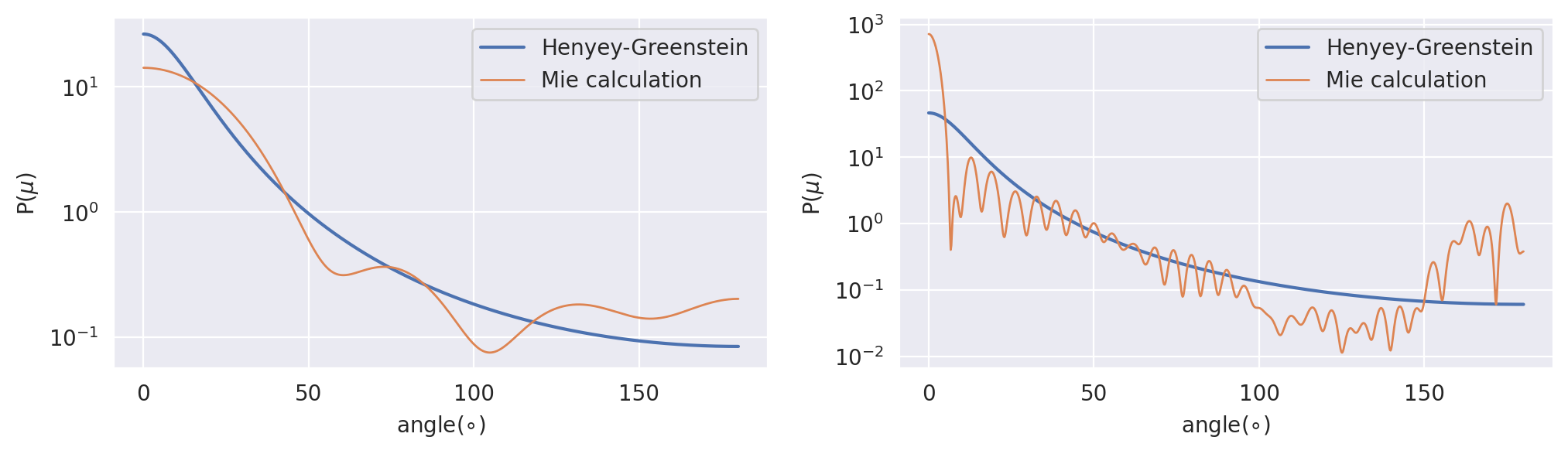}
     \includegraphics[scale =0.35]{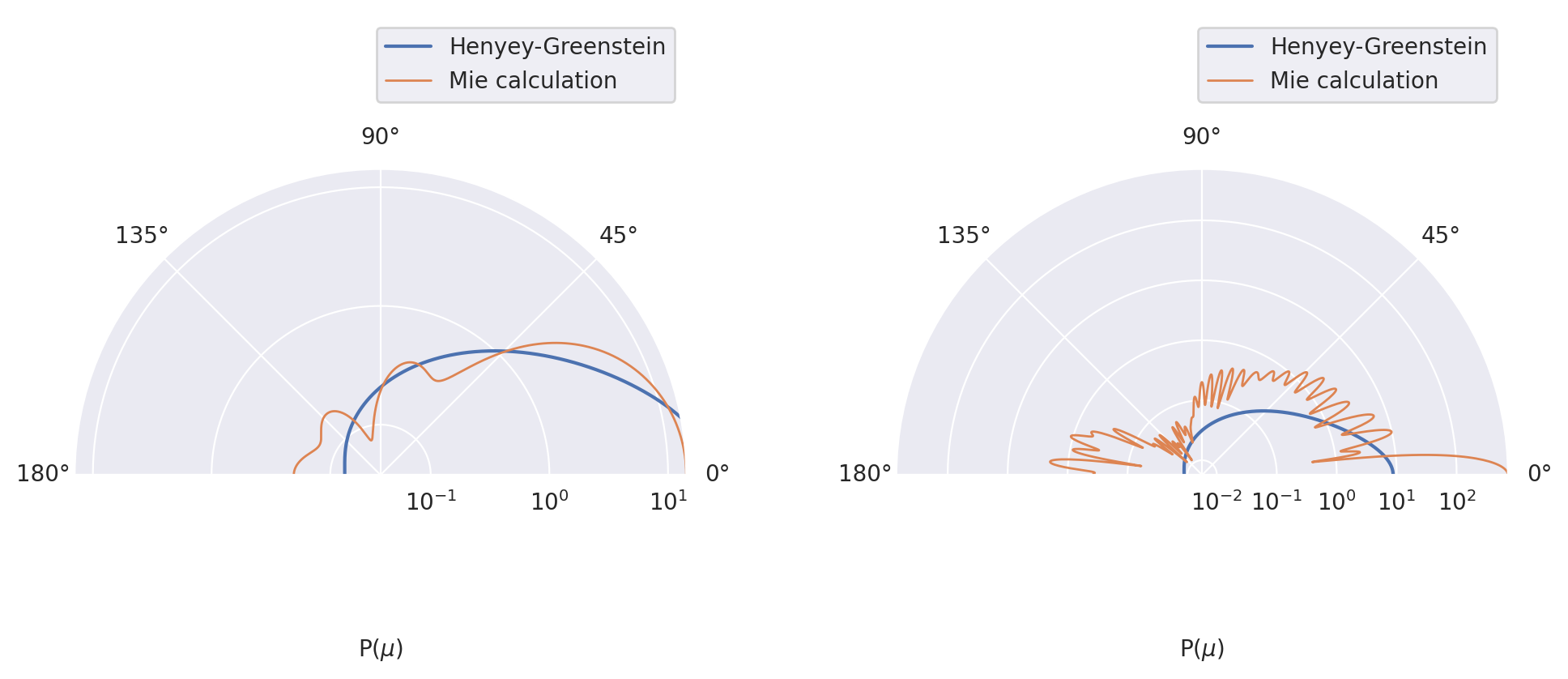}
     \caption[Scattering phase functions]{(Top) The scattering phase functions of particle sizes 0.1 $\mu$m (Left) and (Right)1 $\mu$m. (Bottom) the scattering phase functions as a polar plot. The radial axis shows the angle. }
     \label{fig:opacplot}
 \end{figure}

We have seen how Ly$\alpha$ photons are absorbed, the next phenomenon we take into account is the forward scattering of FUV photons by dust grains. We need to find out if the FUV photons are completely forward scattered or their probability of being forward scattered. The quantity that defines the probability of a photon scattering into a direction $\mu = \mathrm{cos}\theta$ is the scattering phase function, $p(\mu)$. To compute $p(\mu)$ we use Mie calculations and the Henyey-Greenstein phase function computed using the $\texttt{DSHARP}$ package \citep{Birnstiel2018dsharp} with the dust composition used in \citet{Birnstiel2018}. The Henyey-Greenstein phase function is expressed by,
\begin{equation}\label{eq:hg}
   p_g(\mu) = \frac{1}{2} \frac{1-g^2}{(1 +g^2 - 2g\mu)^{3/2}},
\end{equation}
where $g$ is the parameter that tells us how forward peaked a phase function is. It is the amount of forward direction that is retained in a scattering event. This can be computed by,
\begin{equation}\label{eq:anisotropy}
    g  = <\mu> = \int^{1}_{-1} p(\mu)\mu d\mu.
\end{equation}

Typically we consider it to be forward scattering when $\lambda < 2\pi a$, where $a$ is the grain size and $\lambda$, the wavelength of the photon propagation. Fig. \ref{fig:opacplot} shows us the forward peaked nature of the phase functions arrived through Mie Calculation and Henyey-Greenstein function.

Having established the forward scattering nature of the dust we need to find out how much of the flux is scattered forward and how much of it is scattered downward. In order to arrive at this we compute the probability of the photon to be scattered in the forward and downward directions respectively. We use the Henyey-Greenstein function for the same. Now the challenge is defining what a forward scattering direction entails and similarly for a downward scattering direction since physically speaking it is 3 dimensions we are dealing with. Thankfully symmetrical effects help us reduce the constraints. The probability of a photon to be scattered in a direction $\mu \in [\mu_1,\mu_2]$ is given by,
\begin{equation}
    P = \frac{1}{2\pi}\int \int^{\mu_2}_{\mu_1} p(\mu,g) d\mu d\Omega,
\end{equation}
where $p(\mu,g)$ is the Henyey-Greenstein function defined in Eq. \ref{eq:hg}. We can integrate it through the azimuthal direction since the azimuthal direction does not really matter for us, if a photon goes forward but slightly shifted azimuthally it is still forward scattering, this cancels out the $1/2\pi$ pre-factor and we arrive at
\begin{equation}
    P = \int^{\mu_2}_{\mu_1} p(\mu,g) d\mu.
\end{equation}

We define a model where a photon is said to be scattered in the radial direction if $\mu` \in [0,\mathrm{cos}\left(\frac{\pi}{4}\right)]$.  This includes forward and backward scattering. Then, the probability of a photon being scattered in the radial direction is
\begin{equation}
    P_r = \int^{\mathrm{cos}\left(\frac{\pi}{4}\right)}_0 p(\mu.g) d\mu.
\end{equation}
As mentioned before this includes both forward and backward scattering. But if we are interested only in the forward scattering event, then the probability of a forward scattering event is $P_r/2$. Therefore the flux transport in the forward direction is given by
\begin{equation}\label{eq:FUVrad}
    F_{\mathrm{FUV,r}} = \frac{1}{2}P_r F_\mathrm{FUV}e^{-\tau_r},
\end{equation}
where $F_{\mathrm{FUV}} = 0.2F_{\odot,\mathrm{FUV}}$, is the unimpeded FUV flux, $\tau_r$ is the radial optical depth given by Eq. \ref{eq:taur}. 

Similarly we define scattering in the vertical direction if $\mu \in [\mathrm{cos}\left(\frac{\pi}{4}\right),\mathrm{cos}\left(\frac{3\pi}{4}\right)]$. We get
\begin{equation}
    P_z = \int^{\mathrm{cos}\left(\frac{3\pi}{4}\right)}_{\mathrm{cos}\left(\frac{\pi}{4}\right)} p(\mu.g) d\mu.
\end{equation}
This includes both upward and downward scattering. We are interested only in a downward scattering event for which the probability is $P_z/2$. With this computed, we can arrive at the fraction of flux transported vertically downward,
\begin{equation} \label{eq:FUVvert}
    F_{\mathrm{FUV},z} = \frac{1}{2e}P_ze^{-(\tau_z - \Phi)},
\end{equation}
where $\Phi$ is the flaring angle of the disk , and $\tau_z$ the vertical optical depth. Combining Eq. \ref{eq:FUVrad} and Eq. \ref{eq:FUVvert} we get the total FUV flux transported as
\begin{equation}
    F'_{\mathrm{FUV}} = \frac{1}{2}\left(P_r F_\mathrm{FUV}e^{-\tau_r} + \frac{1}{e}P_z F_\mathrm{FUV} e^{-(\tau_z - \Phi)}\right).
\end{equation}

\begin{figure*}
    \centering
    \includegraphics[scale=0.5]{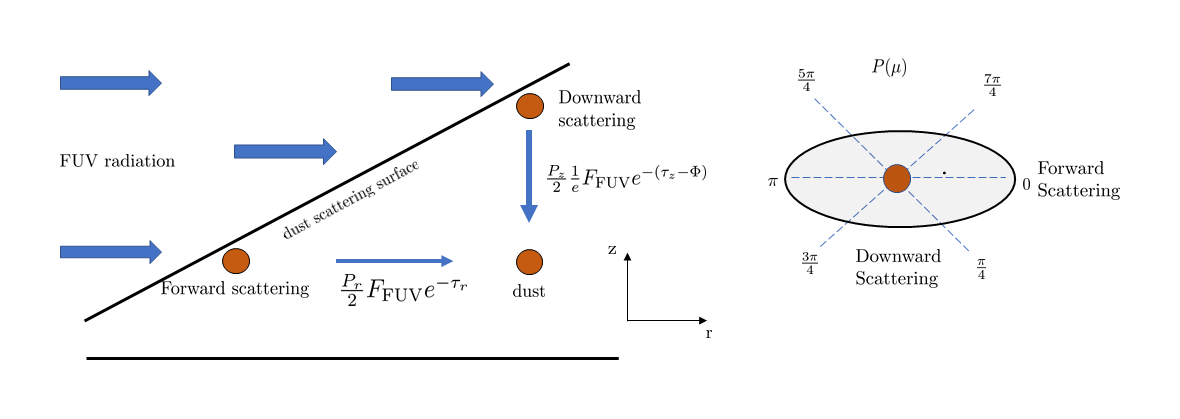}
    \caption[FUV flux scattering model]{(Left)A toy model sketch of a scattering scenario that includes forward and downward scattering (Right) The domains that are defined as forward and downward scattering in the model}
    \label{fig:FUVmodel}
\end{figure*}

Fig. \ref{fig:FUVmodel} shows a broad cartoon sketch of our model of FUV flux transport. The factor $\tau_z-\Phi$ arises from the fact that the scattering happens at the surface $\tau_r = 1$ or $\tau_z = \Phi$ (Eq. \ref{eq:taur}) and we are calculating the vertical transport from $\tau_z = \Phi$ to higher $\tau_z$. $\tau_z$ can be calculated as
\begin{equation} \label{eq:tauz}
    \tau_z = - \int_\infty^z \kappa_{\mathrm{tot}}\rho_d dz,
\end{equation}
where $\rho_d$ is the dust density and $\kappa_{\mathrm{tot}}$ is the total opacity that includes absorption and scattering processes which can be expressed as
\begin{equation}
    \kappa_{\mathrm{tot}} = (1-g) \kappa_{\mathrm{scat}} + \kappa_{\mathrm{abs}},
\end{equation}
where $g$ is the anistropy factor as seen in Eq. \ref{eq:anisotropy}. $\kappa_{\mathrm{scat}}$ and $\kappa_{\mathrm{abs}}$ are the scattering and absorption opacities respectively. We calculate these values using the $\texttt{DSHARP}$ package \citep{Birnstiel2018,Birnstiel2018dsharp} which does Mie calculations to arrive at the opacities. For the calculations, the opacity tables used are from \citet{Segelstein1981}. For larger grain sizes the geometrical effects dominate, so instead of doing the Mie calculations we calculate the geometrical opacity given by
\begin{equation}
    \kappa_{\mathrm{geo}} = \frac{\sigma_{\mathrm{geo}}}{m},
\end{equation}
where $\sigma_{\mathrm{geo}} = \pi a^2$ is the geometric cross section of the dust grain and $m$ is the mass of the dust grain. Now the only piece remaining to calculate the optical depth $\tau_z$ as per Eq. \ref{eq:tauz} is dust density $\rho_d$ which is done by setting up an analytical dust distribution. More details of the dust setup are presented in Appendix \ref{sec:dustdistmath}.

\subsection{Comparison with \texttt{DALI}}\label{subsec-DALI}

Any analytical model requires benchmarking with a physical setup to test its robustness. We make use of the thermo-chemical code \texttt{DALI} (\citet{Bruderer2012}, \citet{Bruderer2013}) which does the required radiative transfer calculation for a protoplanetary disk model. The simulation uses an analytical disk surface density prescription as from \citet{Alessio},\citet{Andrews2011} and \citet{Cleeves2013}. It is described in detail in \citet{Miotello2016}. The vertical distribution is defined by a Gaussian with a scale height angle,
\begin{equation}
    h = h_c \left( \frac{R}{R_c} \right)^\psi,
\end{equation}
from which the physical scale height can be derived as $H = Rh$. The vertical dust model is implemented in two populations of grains, small grains (0.005 $\mu$m - 1 $\mu$m) and large grains (0.005 $\mu$m - 1000 $\mu$m). The surface densities of the large grains are $f\Sigma_\mathrm{dust}$ where $f$ is the fraction of the large grain population and the surface densities of the small grains can be obtained by $(1-f)\Sigma_\mathrm{dust}$. The scale height of the small grains is $Rh$ whereas for the large grains, it is defined by $\chi Rh$ where $0 < \chi < 1$. The vertical density distributions for the small and large grains are expressed as
\begin{equation}
\rho^{\mathrm{small}}_{\mathrm{dust}} = \frac{(1-f)\Sigma_{dust}}{\sqrt{2 \pi}Rh} \mathrm{exp} \left[-\frac{1}{2}\frac{z^2}{h^2}\right],
\end{equation}
\begin{equation}
\rho^{\mathrm{large}}_{\mathrm{dust}} = \frac{(f)\Sigma_{dust}}{\sqrt{2 \pi}\chi Rh} \mathrm{exp} \left[-\frac{1}{2}\frac{z^2}{(\chi h)^2}\right].
\end{equation} 
The values for the parameters for the specific run are given in Table \ref{tab:daliparams}.

\begin{table}
    \centering
    \begin{tabular}{cc}
    \hline
       Parameters  & Value  \\
    \hline
    \hline
         $R_c$ & 60 AU\\
         $\Sigma_c$ & 0.65 g/cm$^2$ \\
         flaring index $\chi$ & 0.2\\
         scale height angle $h_c$ & 0.1\\
         fraction of large grains f & 0.9\\
         flaring index $\psi$ & 0.1 \\
    \hline
    \end{tabular}
    \caption[DALI parameters]{Values for the parameters used in the DALI simulation disk model}
    \label{tab:daliparams}
\end{table}

The simulation gives us the FUV flux values for every (r,z) and we take the vertical data of the flux at r=1 AU. Comparing the flux values obtained from the simulation and the analytical model described in Sect. \ref{subsec:lya} and \ref{subsec:fuvscat} directly as a function of vertical height z would be comparing apples and oranges. This is because the disk structure is different for both the models, resulting in different densities at a given vertical height z. To overcome this, a useful quantity that can be used independent of the model is the optical depth. We use the vertical optical depth $\tau_z$. From eq. \ref{eq:lytau} we can see that optical depth is computed by weighing the vertical height z with the opacity and the density of the particles at the vertical height z. This will get us a model-independent quantity that is used to compare the two models and also to feed into the $\texttt{1DMC}$ code as described in Sec \ref{sec:1DMC}. 

\begin{figure}
    \centering
    \includegraphics[width=0.9\columnwidth]{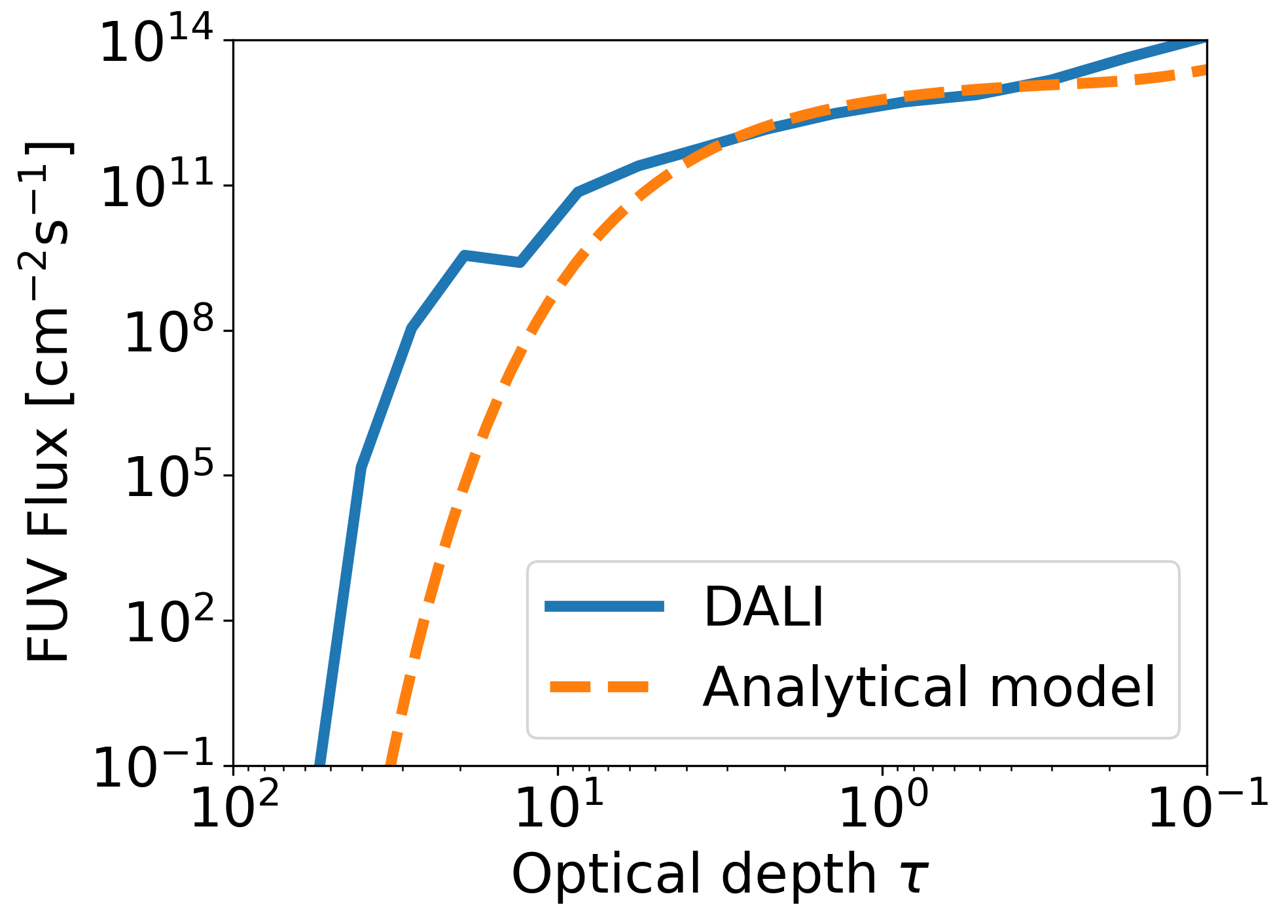}
    \caption[FUV flux of DALI and analytical model]{FUV flux comparison of the DALI simulation and the analytical model. It can be seen that both the models agree around $\tau \sim 1$ where most of the carbon removal occurs}
    \label{fig:flux}
\end{figure}

Fig. \ref{fig:flux} shows the FUV flux as a function of optical depth for both the Analytical Model and the DALI simulation. It can be seen that there is very good agreement at around the $\tau_z = 1$ optical depth. This is where most of the carbon removal occurs since the flux values are the highest there and a large number of photons would interact with the dust to remove carbon. The results of the DALI code show a deeper penetration of the flux when compared with the flux derived from the analytical model. This means that the FUV flux from the DALI simulation is likely to deplete more carbon as the dust densities increase as we go deeper. This arises from the fact that our model does not completely model the scattering of FUV and Ly$\alpha$ photons to the level of a radiative transfer code i.e. DALI which is a code that solves the radiative transfer equation with the Monte Carlo approach while also including much more detailed scattering processes to calculate the temperature and the radiation field \citep{Bruderer2013}. The model we have developed acts as a very good lower limit of the flux values and is almost close to exact at optical depths $\tau_z \approx 1$. This allows us to play around with the stellar parameters to get a physical lower limit, arrive at a physical intuition, and finally compare it with simulation, saving a lot of computational time and arriving at precise results.

\section{Radial optical depth calculation of dust and gas} \label{sec:radoptdepth}
To find the radial optical depth of we follow the method from \citet{Watanabe2008}. For gas
\begin{equation} \label{eq:radintegral}
    \tau_r = \int^r_{R_\star} \kappa_s(T_\star) \rho_g(r',\zeta r')(1+\zeta^2)^{1/2}dr',
\end{equation}
where $\zeta = z/r$,
\begin{equation}
\rho_g(r,z) = \frac{\Sigma_g(r)}{\sqrt{2\pi}h(r)}\mathrm{exp}\left(-\frac{z^2}{2h^2}\right),
\end{equation}
and $\kappa_s = \frac{\sigma}{m_H}$, we take $\sigma = 10^{-21} \mathrm{cm}^{2}$ to consider the H-Layer for the Ly-$\alpha$ scattering(\citet{Bethell2011}), also $\kappa_s = \frac{2 \tau_\nu}{\Sigma_g}$ . The surface density is a radial power law given by
\begin{equation}
    \Sigma_g = \Sigma_{g0} \left(\frac{r}{r_0}\right)^{-1}.
\end{equation}
Plugging these values we arrive at
\begin{equation} \label{eq:radialopt}
\sqrt{\frac{2}{\pi}}\tau_\nu(1 + \zeta^2)^{1/2}\int_{R_*}^{r} \frac{r_0}{r'h(r')} e^{-\left(\frac{\zeta r'}{h(r')}\right)^2/2} dr'
\end{equation}
and to find the scattering surface,
\begin{equation}
    \tau_r = 1.
\end{equation}
The integral in Eq. \ref{eq:radialopt} is numerically computed and solved for $\tau_r = 1$ to get the vertical height z. Now we find the vertical optical depth $\tau_z$ at the height where $\tau_r = 1$ and we find $\tau_z = 0.03 = \Phi$, where $\Phi$ is the flaring angle, this verifies the relation
\begin{equation}
    \Phi \approx \frac{\tau_z}{\tau_r}.
\end{equation}
Now for dust, we get the same relation when $\rho_g$ is replaced with $\rho_d$, the dust density in Eq. \ref{eq:radintegral}. The dust density  $\rho_d = \epsilon \rho_g$, where $\epsilon$ is the dust to gas ratio. 
To calculate the mean opacities for the dust we use
\begin{equation}
    \kappa = \frac{ \int^{a_{\mathrm{max}}}_{a_{\mathrm{min}}}\kappa_{\nu}(a)a^3n(a)da}{ \int^{a_{\mathrm{max}}}_{a_{\mathrm{min}}}a^3n(a)da}, 
\end{equation}
where $\kappa = \frac{3}{4\rho_s a}$ and the number density for each size bin is given by, $n(a) \sim a^{-3.5}$ which is the MRN distribution \citep{Mathis1977}.

\section{Deriving the Analytical dust distribution}\label{sec:dustdistmath}

We start with the advection-diffusion equation \citep{Armitage2020, Fromang2009, Dubrulle1995}:
\begin{equation}
    \frac{\partial \rho_g}{\partial t} - \frac{\partial}{\partial z}\left(D_d \rho_g \frac{\partial}{\partial z}\left(\frac{\rho_d}{\rho_g}\right)\right) + \frac{\partial}{\partial z}\left(\rho_d v_z\right) = 0,
\end{equation}
where $v_z$ is the settling velocity and $D_d$ is the Diffusion constant. Since the model is steady state the temporal derivative vanishes resulting in
\begin{equation}
    \rho_d v_z - D_d \rho_g \frac{\partial}{\partial_z}\left(\frac{\rho_d}{\rho_g}\right) = 0.
\end{equation}
The equation has a solution of the form
\begin{equation}\label{eq:rhodsol}
\rho_d = \rho_g\frac{\rho_{d0}}{\rho_{g0}} \mathrm{exp} \left(\int^z_0 \frac{v_z}{D_d} dz'\right),    
\end{equation}
where $\rho_{g0}, \rho_{d0}$ are the midplane densities of gas and dust respectively. Following \citet{Youdin2007}, we have
\begin{equation}
    D_d = \frac{D_g}{\mathrm{Sc}} = \frac{\alpha c_s H}{1 + \mathrm{St}^2}
\end{equation}
and the settling velocity $v_z$ is given by
\begin{equation}
    v_z = -\Omega^2 t_{\mathrm{fric}}z = - \mathrm{St} \Omega z.
\end{equation}
We also have the Stokes number as
\begin{equation} \label{eq:stokesrho}
    \mathrm{St} = \mathrm{St}_0 \left(\frac{\rho_g}{\rho_{g0}}\right)^{-1},
\end{equation}
where we can get the stokes number in the midplane as
\begin{equation}
    \mathrm{St}_0 = \sqrt{\frac{\pi}{8}}\frac{a\rho_s\Omega}{\rho_{g0}c_s}
\end{equation}
assuming the Epstein regime. 
Plugging the above expressions in the integral in Eq. \ref{eq:rhodsol} we get
\begin{equation}
    \int_0^z \frac{v_z}{D_d} dz' = - \int_0^z \frac{\mathrm{St} \Omega z}{\alpha c_s H} (1 + \mathrm{St}^2).
\end{equation}
Applying Eq. \ref{eq:stokesrho} and integrating we get
\begin{multline}\label{eq:rhozintegral}
    \rho_d(z,a) = \rho_g \frac{\rho_{d0}}{\rho_{g0}} \mathrm{exp} \Biggl[-\frac{\mathrm{St}_0}{\alpha} \left(\mathrm{exp}\left(\frac{z^2}{2H^2} \right)-1\right)\\
    \left(1+\frac{\mathrm{St}^2_0}{2} \left(\mathrm{exp}\left(\frac{z^2}{2H^2}\right)-1\right)\right)\Biggr]
\end{multline}
Now to transform the above vertical dust density distribution into a fragmentation-limited distribution, we follow the steady state distribution derived in \citet{Birnstiel2018} which is a simplified version of \citet{Birnstiel2011a}.

The steady-state surface density distribution is
\begin{equation}\label{eq:steadystatedustdist}
    \sigma(a) = 
    \begin{cases}
     N \left(\frac{a}{a_0}\right)^{0.5} & a_0 <a <a_1 \\
     N \left(\frac{a_1}{a_0}\right)^{0.5} \left(\frac{a}{a_1}\right)^{-0.75}  & a_1\le a\le a_{\mathrm{frag}}
    \end{cases}
\end{equation}
where $a_0, a_1,$ are particle size limits set according to \citet{Birnstiel2018}, $a_\mathrm{frag}$ is the fragmentation size barrier as seen in Eq. \ref{eq:fragbarrier} and N is the normalisation factor which we find out by utilizing
\begin{equation}
    \Sigma_d = \int \sigma(a)da.
\end{equation}
Plugging in Eq. \ref{eq:steadystatedustdist} and solving the integral we get the normalisation factor N as
\begin{equation}
    N =  \frac{\Sigma_d \sqrt{a_0}}{\frac{2}{3}(a_1^{1.5}-a_0^{1.5}) + 4a_1^{5/4}(a_2^{0.25} - a_1^{0.25}) }.
\end{equation}

We know how the dust density varies vertically from Eq. \ref{eq:rhozintegral}. To apply the fragmentation-limited dust distribution to the density, we limit the number density using the steady state surface density distribution from Eq. \ref{eq:steadystatedustdist} and we use a normalisation factor $F_a$ which is an integral of Eq. \ref{eq:rhozintegral} over vertical height $z$. The final fragmentation limited dust density distribution becomes
\begin{multline}
    \rho_d(z,a) = \frac{\sigma(a)}{2F_a} \mathrm{exp} \Biggl[\left(\frac{-z^2}{2H^2}\right)-\frac{\mathrm{St}_0}{\alpha} \left(\mathrm{exp}\left(\frac{z^2}{2H^2} \right)-1\right)\\ \left(1+\frac{\mathrm{St}^2_0}{2} \left(\mathrm{exp}\left(\frac{z^2}{2H^2}\right)-1\right)\right)\Biggr]
\end{multline}

\end{appendix}
\end{document}